\documentclass{aa}
\usepackage{graphicx}
\usepackage[varg]{txfonts}
\usepackage{natbib}
\usepackage{soul}
\newlength{\linwx}
\begin{document}
\title{Influence of the water content in protoplanetary discs \\ on planet migration and formation}
\author{
Bertram Bitsch \inst{1},
\and
Anders Johansen \inst{1}
}
\offprints{B. Bitsch,\\ \email{bert@astro.lu.se}}
\institute{
Lund Observatory, Department of Astronomy and Theoretical Physics, Lund University, 22100 Lund, Sweden
}
\abstract{
The temperature and density profiles of protoplanetary discs depend crucially on the mass fraction of micrometre-sized dust grains and on their chemical composition. A larger abundance of micrometre-sized grains leads to an overall heating of the disc, so that the water ice line moves further away from the star. An increase in the water fraction inside the disc, maintaining a fixed dust abundance, increases the temperature in the icy regions of the disc and lowers the temperature in the inner regions. Discs with a larger silicate fraction have the opposite effect. Here we explore the consequence of the dust composition and abundance for the formation and migration of planets. We find that discs with low water content can only sustain outwards migration for planets up to 4 Earth masses, while outwards migration in discs with a larger water content persists up to 8 Earth masses in the late stages of the disc evolution. Icy planetary cores that do not reach run-away gas accretion can thus migrate to orbits close to the host star if the water abundance is low. Our results imply that hot and warm super-Earths found in exoplanet surveys could have formed beyond the ice line and thus contain a significant fraction in water. These water-rich super-Earths should orbit primarily around stars with a low oxygen abundance, where a low oxygen abundance is caused by either a low water-to-silicate ratio or by overall low metallicity.
}
\authorrunning{Bitsch \& Johansen}\titlerunning{Influence of the water content in protoplanetary discs on planet migration and formation}\maketitle

\section{Introduction}
\label{sec:introduction}

The birth environment for planets is the protoplanetary disc that surrounds newly born stars. Planetary cores must form within a few million years, the typical lifetime of protoplanetary discs  \citep{1998ApJ...495..385H, 2001ApJ...553L.153H, 2009AIPC.1158....3M}, in order to attract a gaseous envelope. Planets with gaseous envelope can roughly be divided into three categories. Gas giants have a mass budget that is dominated by the gaseous envelope, while the ice giants residing beyond the ice line are core-dominated and accrete only a minor amount of gas. The third category of planets with gas envelopes consists of hot and warm super-Earths orbiting close to their host star. Super-Earths with masses above 5 ${\rm M}_{\rm Earth}$ have large radii, consistent with a significant gaseous envelope \citep{2014ApJ...783L...6W}. The densities of smaller super-Earths are more similar to silicate rock \citep{2014ApJ...783L...6W}, although the masses of such small planets are hard to measure precisely. Also, photoevaporation could play a significant role in depleting the primordial envelope of small planets in very hot orbits \citep{2012MNRAS.425.2931O}. In contrast, the terrestrial planets in the solar system are believed to have formed after the dissipation of the gaseous disc \citep{2009GeCoA..73.5150K, 2014Natur.508...84J}. Nevertheless, the dominant part of the total mass of a planetary system must be assembled during the gaseous disc phase, as proven by the prevalence of planets with gaseous envelopes.

The protoplanetary disc sets the conditions for the two dominant physical processes that shape the final architecture of planetary systems, namely mass accretion and orbital migration. Dust grains in protoplanetary discs collide and grow to pebble aggregates of mm-cm sizes \citep{2014prpl.conf..339T}. These pebbles are accreted very efficiently by the growing planetary cores \citep{2010A&A...520A..43O, 2012A&A...544A..32L, 2012A&A...546A..18M}. The largest cores perturb the orbits of the smaller planetesimals and quench their ability to accrete pebbles, guaranteeing that the planetary system ends up with a few massive planets rather than a multitude of embryos \citep{2015Natur.524..322L}. In a previous paper we considered detailed models for the growth and migration of planets \citep{2015A&A...582A.112B}, based on radiative transfer models of evolving protoplanetary discs \citep{2015A&A...575A..28B}. Here we extend this model to consider the effect of the chemical composition of the dust on the formation of planetary systems.

The chemical composition of an exoplanet envelope encodes information about the location in the disc where the planet accreted. In \citet{2014ApJ...794L..12M}  the formation pathways of hot Jupiters through disc migration are constrained by the abundances of oxygen and carbon in their envelopes. They find that the known composition of hot Jupiter atmospheres, which predominantly have a low oxygen abundance, indicates a disc-free migration, meaning a change of semi-major axis induced by gravitational perturbation from another planet or stellar companion. The measurements of the envelope composition by transit spectroscopy is nevertheless complicated by the presence of clouds and hazes in the upper envelope \citep{2014ApJ...791L...9M, 2014ApJ...795..166C,2016Natur.529...59S}.

\citet{2015A&A...580L..13S}  considered instead the influence of the host star metallicity on the composition of the orbiting super-Earths. Direct measurement of the chemical abundance pattern of the host star is more precise than the indirect measurement of abundances in planetary envelopes. The data they collected nevertheless does not allow for clear conclusions to be made for the observed systems, but future analysis combining host star abundances, mass and radius measurements and internal structure models of planets can permit a better characterization of exoplanet interiors and thus their formation.

In this paper we take a complementary approach to \citet{2015A&A...580L..13S} and consider the effect of the chemical composition on the formation and migration of planets. We focus on the role of oxygen and silicon, as well as the overall metallicity of the star. Oxygen and silicon are both alpha elements and are thus enriched by stellar nucleosynthesis roughly following the same pattern. Nevertheless, at any given metallicity, significant variations have been observed in the ratio of silicon to oxygen between individual stars \citep{2014A&A...562A..71B}.

The opacity of the protoplanetary disc is dominated by water ice and silicate grains. We assume that these two components are in a 1:1 ratio in our previous papers \citep{2015A&A...575A..28B, 2015A&A...582A.112B}. Here we expand on this approach and probe the influence of different water-to-silicate ratios on the disc structure and on planet formation and migration. As the opacity changes due to the chemical composition, it directly influence the cooling rate and therefore the temperature profile of the disc. The temperature profile in turn affects both the pebble accretion rates, as lower temperature leads to thinner particle mid-plane layers and increased accretion, and planetary migration, as the radial temperature and entropy gradients determines both the magnitude and the sign of type I  migration \citep{2011MNRAS.410..293P}.

We find that the migration of icy planetary cores to the inner regions of the protoplanetary disc is facilitated in discs with a low water content, as this decreases the temperature gradient near the ice line and thus shrinks the parameter space for outwards migration. A similar result is found for discs with an intrinsically low abundance of small dust grains. Our model thus predicts that water-rich super-Earths orbiting close to their host stars should primarily be found in stars that have a low oxygen abundance.

Our work is structured as follows. In section~\ref{sec:methods} we discuss the methods used to calculate the disc structure and explain our used opacity model for discs with different water-to-silicate ratios. In section~\ref{sec:discstructure} the influence of the water-to-silicate ratio on the disc structure and on planet migration is described. The growth of planetary seeds to planets as a function of the water-to-silicate ratio and the total dust content $Z_{\rm dust}$ is discussed in section~\ref{sec:planetgrowth}. We then discuss the influences and implications of the disc structure on planet formation in section~\ref{sec:discuss}. We finally summarize in section~\ref{sec:summary}.

\section{Methods}
\label{sec:methods}

\subsection{Disc evolution}
\label{subsec:disc}

During the gas phase of the disc, micrometre-sized dust grows through several stages to form planetary cores. This growth over many orders of magnitude in size and mass is directly influenced by the structure of the protoplanetary disc. Dust particles grow to pebbles by coagulation and ice condensation \citep{2010A&A...513A..57Z, 2012A&A...539A.148B, 2013A&A...552A.137R} and drift towards the star due to gas drag \citep{1977Ap&SS..51..153W,2008A&A...487L...1B}. The pebbles then concentrate in the turbulent gas and form planetesimals by their mutual gravity, for example through the streaming instability \citep{2005ApJ...620..459Y, 2007ApJ...662..627J}. These planetesimals then accrete pebbles \citep{2012A&A...544A..32L} and form planetary cores, which finally accrete gas in order to form giant planets \citep{1996Icar..124...62P}. During their growth, the planetary cores and planets migrate through the disc due to gravitational interactions with the gas \citep{1997Icar..126..261W, 2006A&A...459L..17P, 2008A&A...487L...9K, 2009A&A...506..971K}. As all these processes are dependent on the underlying disc structure, it is therefore important to model the structure as accurately as possible.

Recent simulations in 1D \citep{2015arXiv150303352B} and 2D \citep{2013A&A...549A.124B, 2014A&A...564A.135B, 2015A&A...575A..28B} pointed out the importance of including both stellar and viscous heating as well as radiative cooling to determine the disc structure in a self-consistent way. The gas in the disc accretes onto the star with an accretion rate $\dot{M}$. This accretion rate changes in time \citep{1998ApJ...495..385H} and thus influences the structure of the disc. The mass flux $\dot{M}$ is defined as
\begin{equation}
\label{eq:Mdotvr}
 \dot{M} = - 2\pi r \Sigma_g v_r \ .
\end{equation}
Here $\Sigma_g$ is the gas surface density and $v_r$ the radial velocity. Following the $\alpha$-viscosity approach of \citet{1973A&A....24..337S} we can write this as
\begin{equation}
 \label{eq:Mdot}
 \dot{M} = 3 \pi \nu \Sigma_g = 3 \pi \alpha H^2 \Omega_K \Sigma_G \ .
\end{equation}
Here $H$ is the height of the disc and $\Omega_K$ the Keplerian frequency and we adopt a value of $\alpha=0.0054$. In our model $\alpha$ is used to determine the heating and not the disc evolution, which is determined through the $\dot{M}$-time relation of \citet{1998ApJ...495..385H}. The opacity plays here a crucial role, because it determines the cooling of the disc. The opacity is discussed in detail in section~\ref{subsec:opacity}.

The magnetorotational instability can drive the viscosity inside the disc \citep{1998RvMP...70....1B}. This instability drives on the ionisation of atoms and molecules by cosmic and X-rays, so the upper layers are much more easily ionised than the mid-plane. This leads to an MRI non-active zone, the so-called dead zone, in the mid-plane of the disc. In discs with constant $\dot{M}$, a change of viscosity has to be compensated for by an equal change in surface density, but 3D simulations have shown that much of the accretion flow can be carried through the upper layers, resulting in a less dramatic change of surface density compared to flat 2D discs \citep{2014A&A...570A..75B}. On top of that, hydrodynamical instabilities such as the baroclinic instability \citep{2003ApJ...582..869K} or the vertical shear instability \citep{2013MNRAS.435.2610N, 2014A&A...572A..77S} can act as a source of turbulence in the weakly ionised disc mid-plane. As a realistic picture of the source of turbulence in accretion discs is still under debate (see e.g. \citealp{Turner2014}), we neglect the effects of a dead zone and assume a constant $\alpha$ throughout the disc in our simulations.

Recent simulations have also revealed that the disc's accretion might be driven by disc winds (see above) in contrast to viscous spreading \citep{2013ApJ...769...76B, 2013ApJ...772...96B,arXiv:1511.06769}. Already existing literature gives first indications of how disc winds act on the surface density profile of the protoplanetary disc and how this influences planet migration and formation \citep{2015A&A...579A..65O, 2015A&A...584L...1O}, but the simulations to constrain disc winds are under constant development. However, we calculate the disc structure only at given accretion rates, which we link to time through observation \citep{1998ApJ...495..385H}, while disc wind models require to simulate the evolution of the disc over several Myr, which is too time demanding for full 2D simulations. Differences in the scale height coming from different disc models have great influence on the underlying planet formation model, as the scale height determines the pebble accretion rate and the pebble isolation mass: the final mass of the planetary core. We discuss this in section~\ref{sec:discuss}.

\subsection{Opacity}
\label{subsec:opacity}

The opacity is the crucial parameter when calculating the disc structure with heating and cooling, because the cooling rate $\vec{F}$ is directly related to the Rosseland mean opacity ($\vec{F} \propto 1/\kappa_{\rm R}$). Additionally, the absorption of stellar photons is related to the stellar opacity $\kappa_\star$. For more details see \citet{2013A&A...549A.124B}.

\begin{figure}
 \centering
 \includegraphics[width=1.0\linwx]{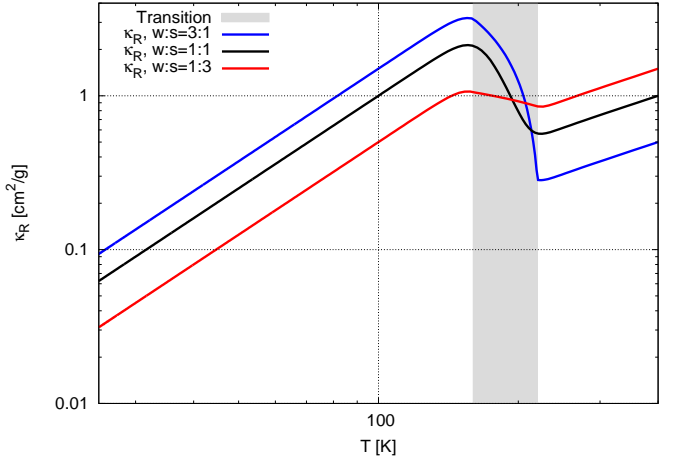}
 \caption{Rosseland mean opacity for micrometre-sized dust grains for the different water-to-silicate ratios. For low temperatures the discs with more water have a larger opacity, because the reflection and absorption parameters for water ice grains yield a higher opacity than for silicate grains.  At the water ice line (grey area) ice grains sublimate, causing a drop in opacity. This drop is larger if the water fraction in the disc is high. The opacity at higher temperatures is larger for discs with more silicates, because more silicate grains are available that can contribute to the opacity. The values of opacity here are calculated for a disc with a total metallicity of $Z_{\rm dust}=0.5\%$ in dust grains.
   \label{fig:kappa}
   }
\end{figure}

The opacity at low temperatures is dominated by micrometre-sized grains. Here we assume that only water ice and silicate grains exist in the protoplanetary disc. However, when the temperature becomes high enough, the water ice grains sublimate, which reduces the opacity inside of the ice line. In Fig.~\ref{fig:kappa} the Rosseland mean opacities for three different water-to-silicate ratios are displayed. The grey region marks the transition region at the water ice line. We investigate here $3$ different compositions of the disc:
\begin{itemize}
 \item water-to-silicate ratio of 3:1
 \item water-to-silicate ratio of 1:1
 \item water-to-silicate ratio of 1:3
\end{itemize}
Changing the water-to-silicate ratio in a disc with a constant abundance of heavy elements results therefore in a change of the total abundance of silicate and water by the corresponding factors. We assume that the contributions to the disc's opacity are only by water and silicates. So, if the water-to-silicate ratio is 3:1, this means the disc has $75\%$ of its heavy elements in water and $25\%$ in silicate. For a water-to-silicate ratio of 1:3, the disc has $25\%$ of its heavy elements in water and $75\%$ in silicates. The case of the water-to-silicate ratio of 1:1 was studied for the disc structure in \citet{2015A&A...575A..28B} and its influence on planet formation in \citet{2015A&A...582A.112B}. 

Evolution of chemical species in accretion discs is a very complex topic that requires special codes that can deal with reaction rates of molecules and how these reactions are affected by the molecular abundances, e.g. \citet{2011ApJ...731..115H}. In principle, by changing the amount of certain elements in the disc, their chemical reaction efficiencies change, leading to changes in the abundances of the various volatile molecules. This could in principle also change the underlying disc structure. However, our model is much simpler and we do not take chemical reactions between the molecules into account. Therefore the opacity is set purely by the amount of available silicate and water grains.

The opacity is also influenced by the total amount of dust grains inside the disc. We let $\Sigma_{\rm dust}$ denote the surface density of heavy elements that are in $\mu m$ size in condensed form. Therefore, the metallicity $Z_{\rm dust}$ is the ratio $Z_{\rm dust} = \Sigma_{\rm dust} / \Sigma_{\rm G}$, assumed to be independent of $r$ in the disc. Our assumption is that the grains are perfectly coupled to the gas, so the dust-to-gas ratio is the same at every location in the disc. In this paper we consider two different cases with a dust metallicity of $Z_{\rm dust}=0.5\%$ and $Z_{\rm dust}=0.1\%$.

The opacity is calculated using Mie-scattering theory from optical constants from water ice \citep{2008JGRD..11314220W} and silicate dust \citep{1994A&A...292..641J, 1995A&A...300..503D}. The final values of the Rosseland mean opacity are computed with the opacity module of radmc-3d\footnote{\url{http://www.ita.uni-heidelberg.de/~dullemond/software/radmc-3d/}}. The differences in the opacity for different water-to-silicate fractions are then translated into differences in the \citet{1994ApJ...427..987B} opacity law, because the radmc-3d opacity model does not account for gas opacities that dominate at very high temperatures that are reached in the very inner and upper layers of the disc.

The absorption and emission features of water ice and silicate grains are different; water ice grains generate a larger opacity than silicate grains. This is the reason why at low temperatures ($T<160$ K) the opacity is higher in the case of a larger water fraction. More precisely, in the case of a water-to-silicate ratio of 3:1, the opacity is $50\%$ larger than for the 1:1 ratio, which in turn has an opacity that is $50\%$ larger compared to the water-to-silicate ratio of 1:3.

At higher temperatures ($T>220$ K) the water ice grains have sublimated and the opacity drops. The drop in opacity is larger for a larger water content in the disc and the resulting opacity is smaller in this case. At these temperatures the only remaining species is silicate, but it exists in different quantities in the $3$ different set-ups, resulting therefore in different opacities. In the case of a water-to-silicate ratio of 3:1, the opacity is $50\%$ smaller compare to the 1:1 ratio, which in turn has an opacity that is $50\%$ smaller than the water-to-silicate ratio of 1:3. In the transition region, where water ice sublimates ($160$ K $<T<220$ K, grey band in Fig.~\ref{fig:kappa}), the opacity connects the two regimes, where we assume that $50\%$ of the water ice is evaporated when $T=190$ K.

The stellar opacity $\kappa_\star$ in itself is dependent only the temperature of the star, but not of the temperature inside the disc. However, when water ice sublimates also the stellar opacity changes, because less dust is available to absorb or re-emit the stellar photons. The stellar opacity changes at the water ice line by exactly the same factors as the Rosseland mean opacity.

\subsection{Planet growth and migration}
\label{subsec:growth}

In the core accretion scenario, the growth of planets can be roughly divided into three stages: 
\begin{itemize}
 \item[1)] the accretion of a core of at least a few Earth masses of solid material
 \item[2)] subsequent contraction of a gaseous envelope until the mass of the gas envelope $M_{\rm env}$ is as large as the mass of the solid core $M_{\rm core}$ 
 \item[3)] runaway gas accretion onto the planet, which allows the planet to become a gas giant.
\end{itemize}
The building of the planetary core is strongly accelerated by accretion of pebbles onto an already existing planetary seed \citep{2012A&A...544A..32L}. We follow here the planet growth model laid out in detail in \citet{2014A&A...572A.107L}. The used code that incorporates planet growth, planet migration and disc evolution is presented in detail in \citet{2015A&A...582A.112B}.

The starting mass of our planetary seeds is at the pebble transition mass
\begin{equation}
\label{eq:Mtrans}
 M_{\rm t} = \sqrt{\frac{1}{3}} \frac{(\eta v_{\rm K})^3}{G \Omega_{\rm K}} \ ,
\end{equation}
where $G$ is the gravitational constant, $v_{\rm K}= \Omega_{\rm K} r$, and
\begin{equation}
\label{eq:eta}
 \eta = - \frac{1}{2} \left( \frac{H}{r} \right)^2 \frac{\partial \ln P}{\partial \ln r} \ .
\end{equation}
Here, $\partial \ln P / \partial \ln r$ is the radial pressure gradient in the disc. Note that the planetesimals formed by the streaming instability are normally a factor of a few smaller than the pebble transition mass \citep{Johansen2015}.

The efficiency of pebble accretion is determined by the pebble scale height in the disc given by
 \begin{equation}
 \label{eq:Hpebble}
  H_{\rm peb} = H\sqrt{\alpha / \tau_{\rm f}} \ ,
 \end{equation}
 where $\tau_{\rm f}$ is the Stokes number of the particles. If the planetary seed's Hill radius is larger than the pebble scale height, the planetary seed can accrete efficiently in a 2D manner
 \begin{equation}
\label{eq:Mdotpebble}
 \dot{M}_{\rm c, 2D} = 2 \left(\frac{\tau_{\rm f}}{0.1}\right)^{2/3} r_{\rm H} v_{\rm H} \Sigma_{\rm peb} \ ,
\end{equation}
where $r_{\rm H} = r [M_{\rm c} / (3M_\star)]^{1/3}$ is the Hill radius, $v_{\rm H}=\Omega_{\rm K} r_{\rm H}$  the Hill speed, and $\Sigma_{\rm peb}$  the pebble surface density. For Stokes numbers larger than $0.1$, the $(\tau_{\rm f}/0.1)^{2/3}$ term vanishes in eq.~\ref{eq:Mdotpebble}, because the planetary seed cannot accrete particles from outside its Hill radius \citep{2012A&A...544A..32L}. However for small planetary seeds, the Hill sphere can be smaller than the pebble scale height, so that the seed accretes in a slow 3D fashion
\begin{equation}
 \dot{M}_{\rm c, 3D} = \dot{M}_{\rm c, 2D} \left( \frac{\pi (\tau_{\rm f}/0.1)^{1/3} r_{\rm H}}{2 \sqrt{2 \pi} H_{\rm peb}} \right) \ . 
\end{equation}
The transition from 3D to 2D pebble accretion is then reached \citep{Morby2015} when
\begin{equation}
 \label{eq:2D3D}
 \frac{\pi (\tau_{\rm f}/0.1)^{1/3} r_{\rm H}}{2 \sqrt{2 \pi}} > H_{\rm peb} \ .
\end{equation}
Clearly this transition depends on the scale height of the pebbles and thus on the scale height of the gas. Therefore, pebble accretion is slower in the initial planetary growth phase in the outer parts of the disc, because the aspect ratio is larger there. The core can then finally reach its pebble isolation mass \citep{2014A&A...572A..35L}
 \begin{equation}
 \label{eq:Misolation}
  M_{\rm iso} \sim 20  \left( \frac{H/r}{0.05}\right)^3 {\rm M}_{\rm Earth} \ ,
 \end{equation}
where pebble accretion is terminated and the contraction of the gaseous envelope can start. We use for the contraction of the gaseous envelope the semi-analytical model described by \citet{2014ApJ...786...21P}, where the contraction phase becomes shorter for more massive planetary cores. This contraction phase additionally depends on the opacity inside the of the planetary envelope, as this determines the cooling and with this the contraction rate of the envelope. Our nominal opacity in the envelope corresponds to similar values as in \citet{2008Icar..194..368M}. We discuss the influence of a slower contraction rate in appendix.~\ref{ap:envelope}. As soon as $M_{\rm env} = M_{\rm core}$, rapid gas accretion can start.

During the whole growth process, planets migrate through the disc. In the initial state, when the planets are small and do not open gaps in the disc, they migrate in type-I migration. We implement this migration rate following the torque formula from \citet{2011MNRAS.410..293P}, which accounts also for the saturation of the corotation torque. As the planets grow in the disc, they can become massive enough to open a gap, which we determine through the gap opening criteria (with $\Sigma_{\rm Gap} < 0.1 \Sigma_{\rm g}$), when
\begin{equation}
\label{eq:gapopen}
 \mathcal{P} = \frac{3}{4} \frac{H}{r_{\rm H}} + \frac{50}{q \mathcal{R}} \leq 1 \ ,
\end{equation}
where $r_{\rm H}$ is the Hill radius, $q=M_{\rm P} / M_\star$, and $\mathcal{R}$ the Reynolds number given by $\mathcal{R} = r_{\rm P}^2 \Omega_{\rm P} / \nu$ \citep{2006Icar..181..587C}. When the planets open a gap, they transition into type-II migration, in which planets follow the viscous evolution of the protoplanetary disc. However, if the planet is much more massive than the gas outside the gap, it slows down the viscous accretion. This happens if $M_{\rm P} > 4\pi \Sigma_{\rm g} r_{\rm P}^2$, which leads to the migration time scale of
\begin{equation}
\label{eq:typeII}
 \tau_{\rm II} = \tau_{\nu} \times \max \left(1 , \frac{M_{\rm P}}{4\pi \Sigma_{\rm g} r_{\rm P}^2} \right) \ ,
\end{equation}
resulting in slower inward migration for massive planets \citep{2013arXiv1312.4293B}. However, recent studies have challenged the viscous type-II inward migration, showing a dependence on the planetary mass and mass flux \citep{2015A&A...574A..52D}. This might lead to the conclusion that the slowing down effect of the planet's inertia on the disc accretion rate is too strong and the real type-II migration rate should be faster. We discuss this in appendix~\ref{ap:migration}.

\section{Disc structure}
\label{sec:discstructure}

We investigate here not only the influence of the different water-to-silicate ratios on the disc, but also the influence of the total amount of dust $Z_{\rm dust}$ on the disc structure. We therefore study two values of the dust abundance, $Z_{\rm dust}=0.5\%$ and $Z_{\rm dust}=0.1\%$. To calculate the disc structure we use the 3D radiative hydrodynamical code FARGOCA as presented first in \citet{2014MNRAS.440..683L}, using the set up of \citet{2015A&A...575A..28B} in 2D with the opacity modifications presented in section~\ref{subsec:opacity} to incorporate the different water-to-silicate ratios.

\begin{figure}
 \centering
 \includegraphics[width=1.0\linwx]{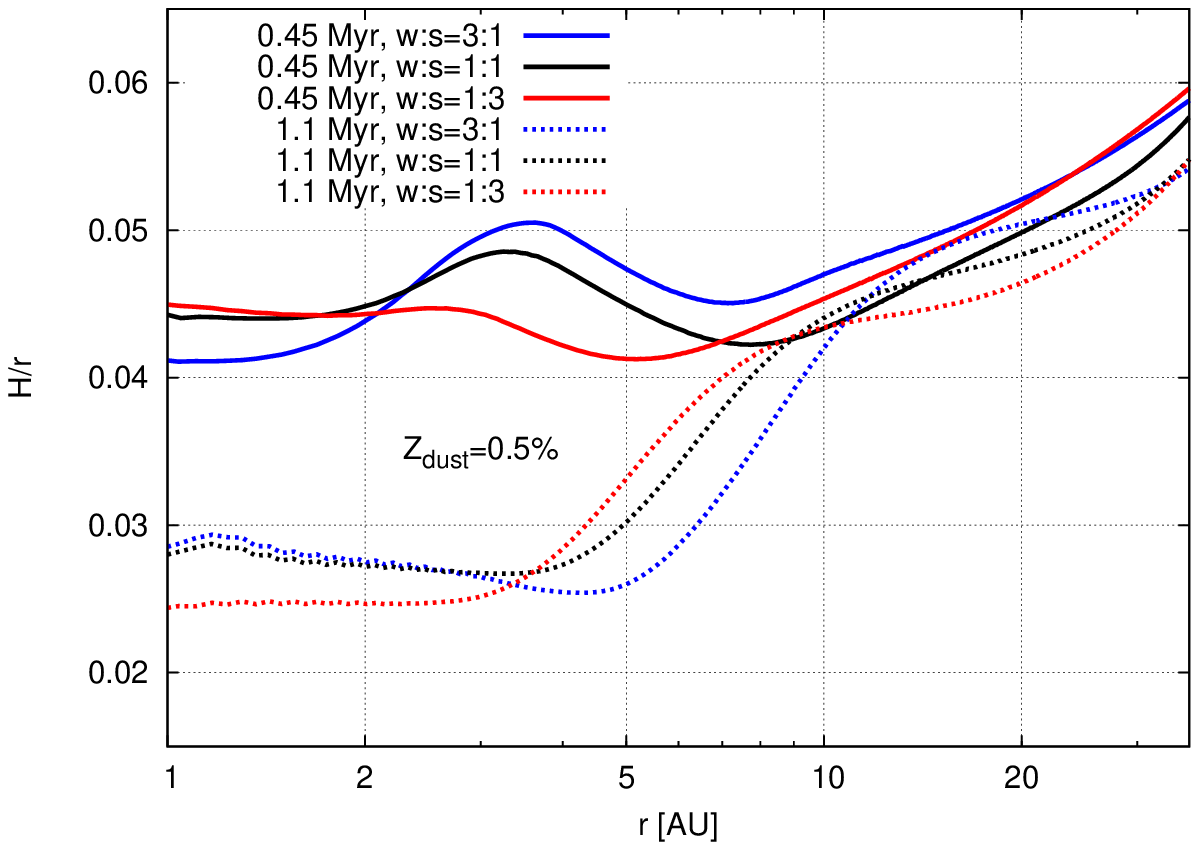}
 \includegraphics[width=1.0\linwx]{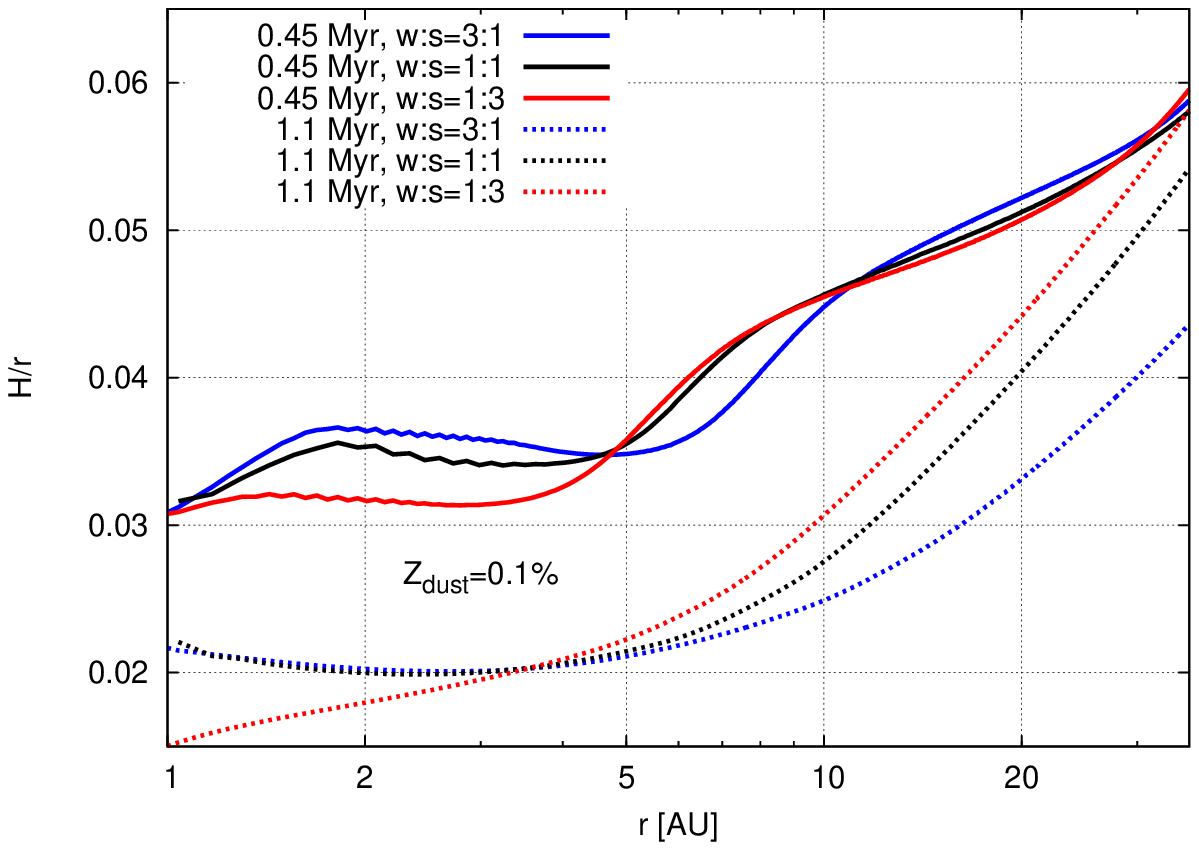}
 \caption{Aspect ratio $H/r$ for discs with different water-to-silicate ratios. The top plot shows $H/r$ for a dust metallicity of $Z_{\rm dust}=0.5\%$ , while the bottom plot has $Z_{\rm dust}=0.1\%$. A larger fraction of water causes a larger bump in the $H/r$ profile at the water ice line, as expected from the opacity transition (see Fig.~\ref{fig:kappa}). At lower dust metallicities the bump at the water ice line is less pronounced compared to higher dust metallicities and resides close to or even inside of $1$ AU at the late stages of the disc evolution. For all water-to-silicate ratios, lower accretion rates $\dot{M}$ result in smaller $H/r$, because the disc is less massive and therefore produces less viscous heating. 
   \label{fig:Hrdisc}
   }
\end{figure}

The change of opacity causes a change in the cooling of the disc. In particular an increase in the Rosseland mean opacity $\kappa_{\rm R}$ results in a decrease in cooling, allowing a higher temperature of the disc. A higher temperature of the disc directly translates to a larger aspect ratio $H/r$. This can be seen in Fig.~\ref{fig:Hrdisc}, most clearly in the top panel in the $\dot{M} =3.5 \times 10^{-8} M_\odot$/yr case, which corresponds to a time evolution of $0.45$ Myr. The larger the water fraction in the disc is, the larger the bump in the aspect ratio at $\sim 3$ AU, which corresponds to the opacity transition at the water ice line $r_{\rm ice}$. 

Interior to $r_{\rm ice}$, the water ice grains have sublimated and only the silicate grains remain, which then dominate the opacity. A lower water-to-silicate ratio therefore results in a higher opacity. This in turn causes again a lower cooling rate, allowing for higher temperatures. This can be seen clearly for the discs with $t=0.45$ Myr in the top panel of Fig.~\ref{fig:Hrdisc}, where the disc with the largest silicate content has the largest aspect ratio in the inner disc ($r<2$ AU).

In the outer parts of the disc, the disc structure is quite similar for all different water-to-silicate ratios. The heating of the outer disc is provided by stellar irradiation, so a higher opacity allows more efficient absorption of stellar irradiation. However, this means that at large orbital distances less stellar heating arrives in discs with high opacity, because it is absorbed at smaller orbital distances, which would result in a lower disc temperature compared to discs with lower opacity. But, the opacity has an opposite effect on cooling. A larger opacity results in a lower cooling rate compared to a small opacity. These two effects seem to balance in the outer regions of the disc for the different water-to-silicate ratios, resulting in similar disc structures.

For the $\dot{M} =8.75 \times 10^{-9} M_\odot$/yr case, which correspond to an evolution time of $1.1$ Myr, the observed trend continues. The disc with the largest water-to-silicate ratio features the largest aspect ratio at the ice line (around $1$ AU). The lower accretion rates results in a reduction of the viscous heating. The consequence is that the water ice line moves closer to the central star, and with it the bump in the aspect ratio. For the outer disc, which is dominated by stellar irradiation, again only minor differences are visible in the structure between the discs with different water-to-silicate ratios. However, as the disc's accretion rate decreases in time, the disc with a water-to-silicate ratio of 3:1 shows a smaller aspect ratio in the outer parts, because the bump at the water ice line shields the outer disc from stellar irradiation efficiently.

The simulations with a smaller dust content of $Z_{\rm dust} = 0.1\%$ (bottom panel in Fig.~\ref{fig:Hrdisc}) show the same trends for the water-to-silicate ratios regarding the aspect ratio, mainly that discs with a larger water fraction have a larger aspect ratio in the inner part of the disc. Also for earlier disc evolution times, the aspect ratio in the outer parts of the disc is very similar for the different water-to-silicate ratios. 

However, at the late stages of the disc evolution clear differences in the aspect ratio arise. As in the case of high $Z_{\rm dust}$, the inner disc is able to shield the outer disc from stellar irradiation. This effect is more efficient in discs with a higher opacity, so in discs with a larger water-to-silicate fraction. This means less stellar irradiation is absorbed by the outer disc, resulting in a cooler disc. At the same time a higher opacity (larger water content in the disc) results in a smaller cooling rate compared to the discs with a lower opacity (low water content). In the discs with high metallicity ($Z_{\rm dust} = 0.5\%$) these effects are in balance, so that the disc structure in the outer disc is the same for all water-to-silicate ratios except for the late stages of the disc evolution. However this balance is apparently broken for low metallicity ($Z_{\rm dust}= 0.1\%$), resulting in a lower aspect ratio for the disc with the larger water content.

As in previous work \citep{2015A&A...575A..28B} we use the disc evolution equation given by \citet{1998ApJ...495..385H} to have a relation between $\dot{M}$ and time. It is given by 
\begin{equation}
\label{eq:harttimenew}
 \log \left( \frac{\dot{M}}{M_\odot /\text{yr}} \right) = -8.00 - 1.40  \log \left( \frac{t_{\rm disc}+10^5\text{yr}}{10^6 \text{yr}} \right) \ ,
\end{equation}
and it allows us to associate a time to a given $\dot{M}$ value. As in \citet{2015A&A...575A..28B} we fit the disc structure by an analytical formula, which allows us to calculate the disc structure at any given orbital distance for a given $\dot{M}$. We then use these fits of the disc structure with eq.~\ref{eq:harttimenew} for our time evolving disc model needed for the planet formation studies presented in section~\ref{sec:planetgrowth}.

\subsection{Influence on planet migration}
\label{subsec:migration}

The changes of the disc structure affect all stages of planet formation. As shown in \citet{2015A&A...582A.112B}, planet migration is of crucial importance, because the planetary seed must form much further away from the star than compared to the final orbital position of the planet. We therefore discuss here the influence of the changed disc structure due to the different water-to-silicate ratios on planet migration.

\begin{figure}
 \centering
 \includegraphics[scale=0.7]{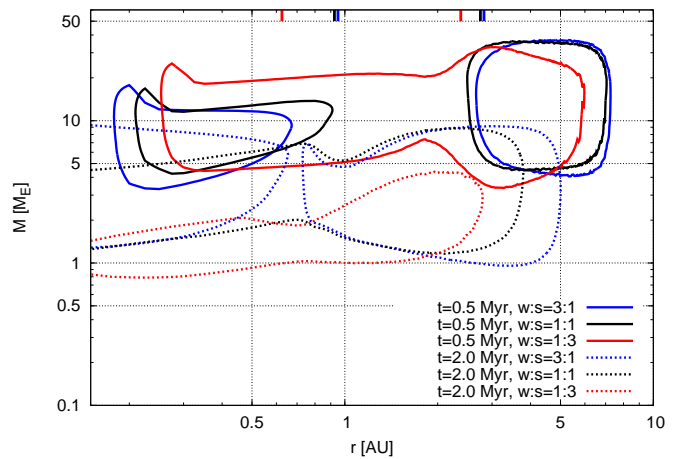}
 \caption{Regions for outward migration in discs with different water-to-silicate ratios ($Z_{\rm dust} = 0.5\%$) and different times for planets in type-I-migration. Planets located inside of the solid lines migrate outwards, while planets outside the lines migrate inwards. As the disc evolves in time, the strong negative gradients in entropy indicated by radial decreases in $H/r$ (Fig.~\ref{fig:Hrdisc}) move towards the star and thus the regions of outward migration move inwards. The small ticks on the top of the plot correspond to the location of the ice line for the different simulations. A clear difference in the regions of outward migration for the different ice to silicate ratios is visible for early times, while the differences become smaller for late evolution times. 
   \label{fig:Migcont}
   }
\end{figure}

Once planetary embryos have formed, they are subject to planetary migration (see \citealp{2013arXiv1312.4293B} for a review). Migration rates can be computed by 2D or 3D simulations \citep{2009A&A...506..971K, 2011A&A...536A..77B}, however, such simulations are quite computationally intensive. We therefore calculate the torque on an embedded planet from the underlying disc structure \citep{2011MNRAS.410..293P}, an approach which matches quite well the results of 3D simulations \citep{2011A&A...536A..77B, Lega2015}. The torque acting on an embedded planet depends crucially on the gradients of surface density and entropy, emphasizing the importance of the disc structure discussed in the previous section. In Fig.~\ref{fig:Migcont} the regions of outward migration in discs with $Z_{\rm dust} = 0.5\%$ and different water-to-silicate ratios for two given accretion rates $\dot{M}$ are displayed. 

At an evolution time of $0.5$ Myr, two regions of outward migration exist except for the disc with a water-to-silicate ratio of 1:3. The outer region of outward migration is originating at the water ice line, while the inner region of outward migration is caused by the silicate evaporation line. When the silicate grains evaporate, the opacity is determined by molecules. In the disc that has the largest silicate fraction, the two regions of outward migration are merged into one big region. The merging of the two regions has its origin in the smaller bump in temperature (and therefore $H/r$) at the water ice line (Fig.~\ref{fig:Hrdisc}), which is too small to cause a clear cut between the two regions of outward migration. For this given accretion rate $\dot{M}$ and time $t$, planets with a minimum of $5$ ${\rm M}_{\rm Earth}$ and a maximum of $\sim 40$ ${\rm M}_{\rm Earth}$ can migrate outwards close to the water ice line, independent of the water-to-silicate ratio.

At an evolutionary time of $2$ Myr, the water ice line is closer to the star and with it the regions of outward migration \citep{2014A&A...564A.135B, 2015A&A...575A..28B}. For the discs with water-to-silicate ratio of 1:1 and 1:3 only one region of outward migration exists. This is caused by the fact that the bump at the water ice line also shrinks as the accretion rate reduces, which then in turn allows the two regions of outward migration to merge together, as also seen at $0.5$ Myr for the largest silicate fraction in the disc. On the other hand, the large water content enhances the bump at the water ice line, separating the two regions of outward migration. Additionally, at low accretion rates the region of outward migration can now contain planets with $1 {\rm M}_{\rm Earth} < M_{\rm P} < 10 {\rm M}_{\rm Earth}$. The changing of the regions of outward migration with time has important consequences for the growth tracks of planets \citep{2015A&A...582A.112B}. We discuss this in detail in section~\ref{sec:planetgrowth}.

When the planets become so massive that they start opening a gap in the disc, their migration rates start to differ from the pure type-I migration rate and the torque can no longer be calculated by using the model of \citet{2011MNRAS.410..293P}. When the planet becomes massive enough to open a gap in the disc, it migrates in type-II migration. In the intermediate mass region, where planets have not fully opened a gap, yet, but only a partly gap, a smoothing function for the migration speed is used to connect these two migration regimes. The detailed methods how we calculate the migration speeds and directions are described in \citet{2015A&A...582A.112B}.

\section{Planet growth}
\label{sec:planetgrowth}

The planet growth and evolution model is described in detail in \citet{2015A&A...582A.112B} and in section~\ref{subsec:growth}. The only difference compared to \citet{2015A&A...582A.112B} is the underlying disc structure that depends on the considered water-to-silicate ratio (section~\ref{sec:discstructure}).

\subsection{High dust content}

We investigate here the influence of the different disc structures caused by the different water-to-silicate ratios on the formation of planets. In this section the underlying metallicity in dust particles in the disc is $Z_{\rm dust} = 0.5\%$. The metallicity of pebbles is fixed to $Z = 1.0\%$.

\subsubsection{Single evolution tracks}

\begin{figure}
 \centering
 \includegraphics[scale=0.7]{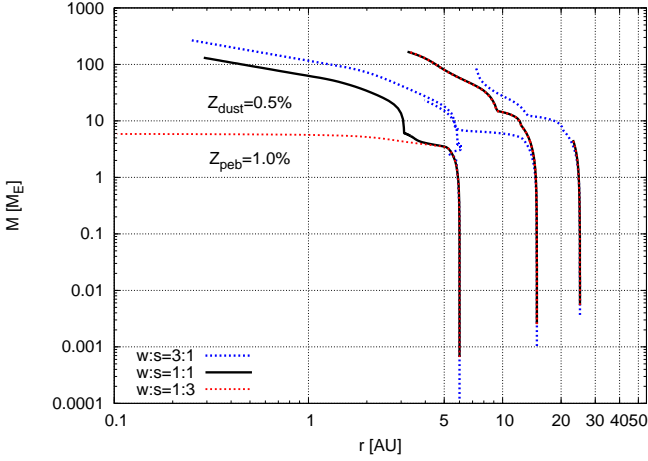}
 \caption{Growth tracks of planets that accrete pebbles in an evolving protoplanetary disc. The planets are placed in discs with different water-to-silicate ratios at a disc evolution time of $t_0=2$ Myr. The simulations are stopped when $t_f=3$ Myr is reached, meaning the planet and disc evolve for $1$ Myr. Note that the different resulting planetary masses and orbital distances are a result of the different underlying disc structures (Fig.~\ref{fig:Hrdisc}) and migration patters (Fig.~\ref{fig:Migcont}).
   \label{fig:Envhockey}
   }
\end{figure}

We follow here the evolution tracks of single planetary seeds. These seeds are put at different orbital distances $r_0$, but they all start in their individual evolution at $t_0=2$ Myr. As the final lifetime of the disc is $3$ Myr in our simulations, each planetary seed is evolved for $1$ Myr. In Fig.~\ref{fig:Envhockey} the evolution tracks for planetary seeds starting at $6$, $15$ and $25$ AU in discs with a water-to-silicate ratio of 3:1, 1:1 and 1:3 are displayed.

The planetary seeds starting at $25$ AU accrete pebbles quite slowly, because the pebble surface density is low in the outer disc. In the disc with a water-to-silicate ratio of 1:1 and 1:3, the accretion is so slow that the planets do not even reach pebble isolation mass. However, in the case of the water-to-silicate ratio of 3:1, the planet reaches pebble isolation mass, contracts a gaseous envelope and enters runaway gas accretion, where its evolution is just stopped at the end of the discs lifetime when the planet reaches $\sim 100 {\rm M}_{\rm Earth}$. The reason why the planet in the water-to-silicate 3:1 disc can reach a much higher planetary mass compared to the planets in the other discs lies in the disc structure.

In the disc with water-to-silicate ratio of 3:1, the aspect ratio far out in the disc at late times ($t>2$ Myr) is a bit smaller than the 1:1 and 1:3 cases. A smaller aspect ratio increases the growth of the planet, because
\begin{itemize}
 \item[i)] the particle sizes are larger, thus allowing a faster accretion
 \item[ii)] the pebble scale height is smaller (eq.~\ref{eq:Hpebble}), allowing a faster transition into the more efficient 2D accretion regime
\end{itemize}
However, the smaller aspect ratio reduces the pebble isolation mass (eq.~\ref{eq:Misolation}), so that planets in the disc with a water-to-silicate ratio of 3:1 have smaller planetary cores. But here the mass of the core is still sufficiently high, allowing the contraction of a gaseous envelope within a few $100$ kyrs, so the planet can evolve into a gas giant. 

During their evolution, the planets starting at $25$ AU that do not reach pebble isolation mass stay in the outer disc where they formed, while the planet formed in the disc with a water-to-silicate ratio of 3:1 migrates down to $\sim 7$ AU, because type-I migration is faster for massive planets.

For the planetary seeds starting at $15$ AU, a similar behaviour can be observed. In the discs with a low water fraction the planetary seeds reach pebble isolation mass, contract their gaseous envelope and then become gas giants, while migrating into the inner disc to $\sim 3$ AU. The planet in the water rich disc reaches pebble isolation mass faster than the counterparts in the water poor discs, for the reasons stated above. But, again, its core is then smaller but still large enough to attract a gaseous envelope and go into runaway gas accretion. Because gas envelope contraction is slower than pebble accretion, it results in a horizontal movement of the planet in the orbital distance - planetary mass diagram (Fig.~\ref{fig:Envhockey}), which is not seen as strongly in the planets in discs with water-to-silicate fraction of 1:1 and 1:3. Note that the planet at that point is too massive to be trapped in a region of outward migration (Fig.~\ref{fig:Migcont}).

The planetary seeds starting at $6$ AU show very different evolution tracks. These differences are caused by small differences in the disc structure. For a water-to-silicate ratio of 1:1 and 1:3, the planet reaches an isolation mass of roughly $3$ Earth masses, while it only reaches $\sim 2$ Earth masses in a disc with water-to-silicate ratio of 3:1. This leads to different contraction times of the gaseous envelope, which takes longer for the seed in the disc with water-to-silicate ratio of 3:1. As the seed grows in this disc its semi major axis evolution follows the evolution of the zero-torque radius until the planet reaches runaway gas accretion and outgrows the region of outward migration. At the end of the disc's lifetime the planet is stranded at $\sim 3.5$ AU and its composition is dominated by the gaseous envelope of roughly $20$ Earth masses.

The seed in the disc with a water-to-silicate ratio of 1:1 reaches a higher pebble isolation mass and can thus contract its envelope faster. During this evolution the planet is trapped in the region of outward migration that stops its inward motion towards the star. As the planet reaches runaway gas accretion it outgrows the region of outward migration and moves inwards as it grows. It is then stranded at $\sim 0.3$ AU with roughly the size of Saturn. The seed in the disc with a water-to-silicate ratio of 1:3 has a similar size and could thus accrete a gaseous envelope in a similar time, however as the region of outward migration can only stop migration of smaller planets (see Fig.~\ref{fig:Migcont}) it drifts into the inner disc as a super-Earth planet and hits the inner edge of the disc.

All these differences in final orbital position and planetary masses are a direct consequence of the difference in the disc structure, even though these differences are actually quite small (Fig.~\ref{fig:Hrdisc}). As the disc evolves in time, we probe in the next section different initial times $t_0$ when the planetary seed is placed in the disc.

\subsubsection{Growth maps}

The growth tracks shown in Fig.~\ref{fig:Envhockey} are for individual planets released at given orbital distances $r_0$ and at an initial time $t_0$. Here we investigate the parameter space for planets starting with $3$ AU $<r_0<50$ AU and with $100$ kyr $<t_0<3$ Myr. The whole disc lifetime is fixed to $3$ Myr, meaning that planets that are starting at a later time $t_0$, only have a shorter time to grow and migrate through the disc compared to planets that start at a smaller $t_0$.

\begin{figure}
 \centering
 \includegraphics[scale=0.7]{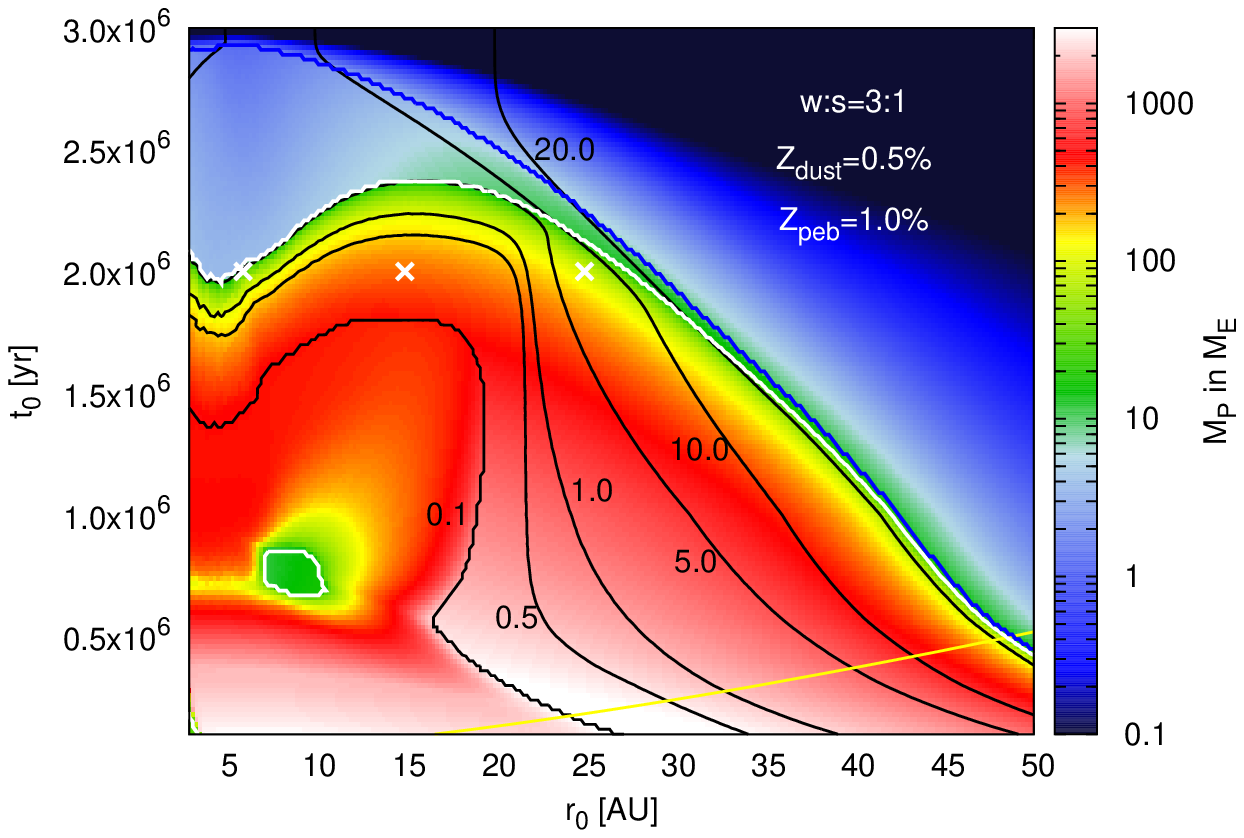}
 \includegraphics[scale=0.7]{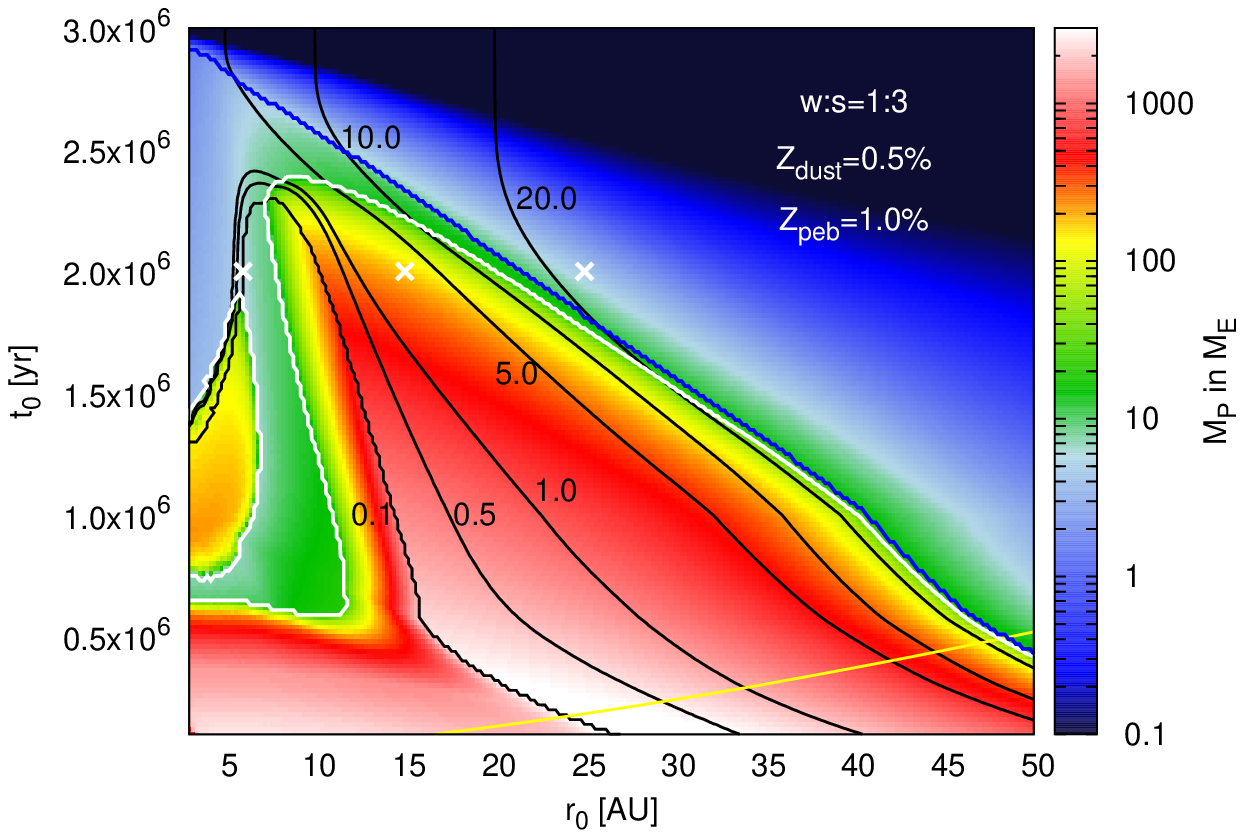}
 \caption{Final masses of planets ($M_{\rm P} = M_{\rm c}+M_{\rm env}$) as a function of initial radius $r_{\rm 0}$ and initial time $t_{\rm 0}$ in discs with a dust metallicity of  $Z_{\rm dust}=0.5\%$. The top plot features an water-to-silicate ratio of 3:1, while the bottom plot features a water-to-silicate ratio of 1:3. Planets that are below the dark blue line have reached pebble isolation mass and can accrete gas. All planets that are inside the white lines have $M_{\rm c}<M_{\rm env}$, indicating that they have underwent runaway gas accretion. The yellow line marks the pebble production line; planets below the yellow line can not have formed by pebble accretion, because the pebbles have not yet formed at the insertion time of the planet, see \citet{2014A&A...572A.107L}. The black lines indicate the final orbital distance $r_{\rm f}$ of the planet. The white crosses indicate $r_0$ and $t_0$ for the evolution tracks shown in Fig.~\ref{fig:Envhockey}. 
   \label{fig:Z010M05}
   }
\end{figure}

In Fig.~\ref{fig:Z010M05} we show the final planetary masses and orbital distances of planetary seeds that are put into a disc at a starting time $t_0$ and a starting orbital distance $r_0$. Each planetary seed at a given $r_0$ and $t_0$ value undergoes an evolution track like the ones shown in Fig.~\ref{fig:Envhockey}. For example, the growth tracks shown in Fig.~\ref{fig:Envhockey} have their $r_0$ and $t_0$ at the location of the white crosses in Fig.~\ref{fig:Z010M05}. The top panel in Fig.~\ref{fig:Z010M05} features a water-to-silicate ratio of 3:1, while the bottom panel features a water-to-silicate ratio of 1:3. We do not display the $r_0$-$t_0$ map for planets evolving in the disc with a water-to-silicate ratio of 1:1, because this is extensively discussed in \citet{2015A&A...582A.112B}, see their Fig.4.

The general planet formation picture is quite similar for all discs, no matter their water-to-silicate ratio:
\begin{itemize}
 \item pebble accretion is efficient enough that planets that form early grow all the way to become gas giants, even at large orbital distances
 \item gas giant planets experience a significant reduction of their semi-major axis due to planet migration
 \item planetary seeds emerging at the late stages of disc evolution form also smaller planets (like super-Earths, ice giants and ice planets)
\end{itemize}
However, there are a few differences between the two extremes of high and low water content in the disc.

In the case of a high water content (3:1 water-to-silicate ratio) giant planets can form at larger distances and later times compared to the case of low water content (1:3 water-to-silicate ratio). This is caused by the slightly lower $H/r$ in the outer disc, which reduces the particle scale height and therefore allows faster pebble accretion (see the $25$ AU growth track in Fig.~\ref{fig:Envhockey}). In the late stages of the disc evolution ($t>2$ Myr), the formed planets are small in the inner disc (low core mass leads to a long contraction time of the envelope) and follow the zero torque migration radius, when they reach it in the inner disc. In the case of a higher water content, this zero torque radius is farther out compared to a disc with low water content. Planetary seeds that form with $t_0<2$ Myr in the disc with high water content end up as gas giants.

However, if planets form in the disc with low water content, the formation of small planets (up to a few Earth masses and $r_f<10$ AU) is possible at nearly all initial times with $t_0>700$ kyr. Interestingly in this case, these low mass planets can even migrate all the way to the inner disc ($r_f<1$ AU), which is not possible in the discs with high water content, where they are trapped in the region of outward migration. Inside the disc with low water content planets can outgrow the zero migration region at a lower planetary mass, which allows them to migrate into the inner disc, before they reach $M_{\rm env} \sim M_{\rm core}$ and runaway gas accretion starts. However, note that we stop our planet evolution model as soon as the planet hits the inner edge of the disc at $0.1$ AU, because planet growth (especially via gas accretion) is unknown in these regions of the disc. 

The small difference in the disc structure caused by the different water-to-silicate ratios (see Fig.~\ref{fig:Hrdisc}) has important consequences for the formation of planets. A low water-to-silicate ratio allows low mass planets to migrate into the inner disc before they can attract a gaseous envelope and grow to gas giants.

\subsection{Low dust content}

The thermodynamical structure of the protoplanetary disc is not only determined by the water-to-silicate ratio, but also by the amount of dust grains inside the disc (see Fig.~\ref{fig:Hrdisc}). In principle the amount of dust grains in the disc can change in time, for example due to grain growth. However, larger grains do not contribute to the opacity for most realistic size distributions, so the opacity is reduced when grains grow. On the other hand, the amount of micrometre-sized dust can increase, for example through destructive collisions of larger dust aggregates and pebbles. As shown in Fig.~\ref{fig:Hrdisc}, a smaller amount of dust grains results in a colder disc, because the cooling rate increases \citep{2015A&A...575A..28B}. We now investigate how planet formation is influenced if the disc has a metallicity of dust particles of $Z_{\rm dust}=0.1\%$. In particular we also investigate here the influence of the water-to-silicate ratio.

\begin{figure}
 \centering
 \includegraphics[scale=0.7]{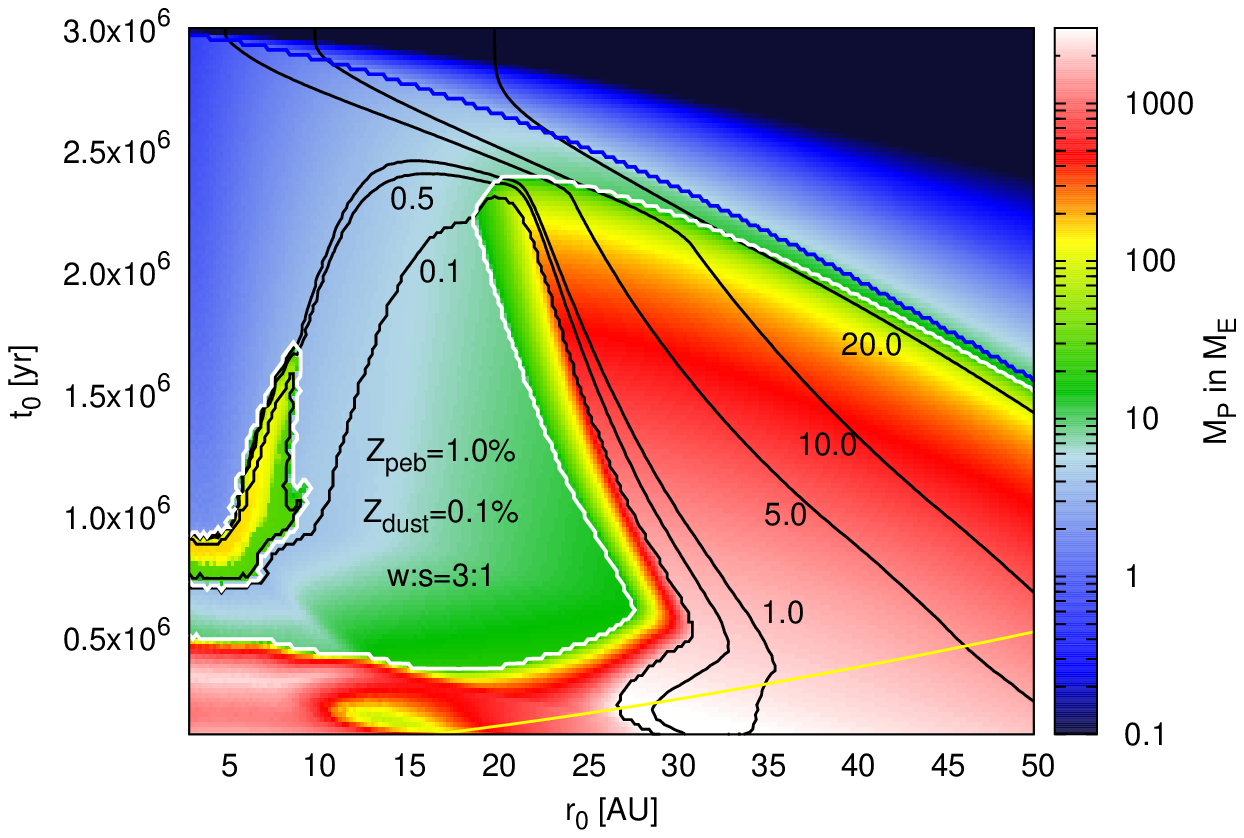}
 \includegraphics[scale=0.7]{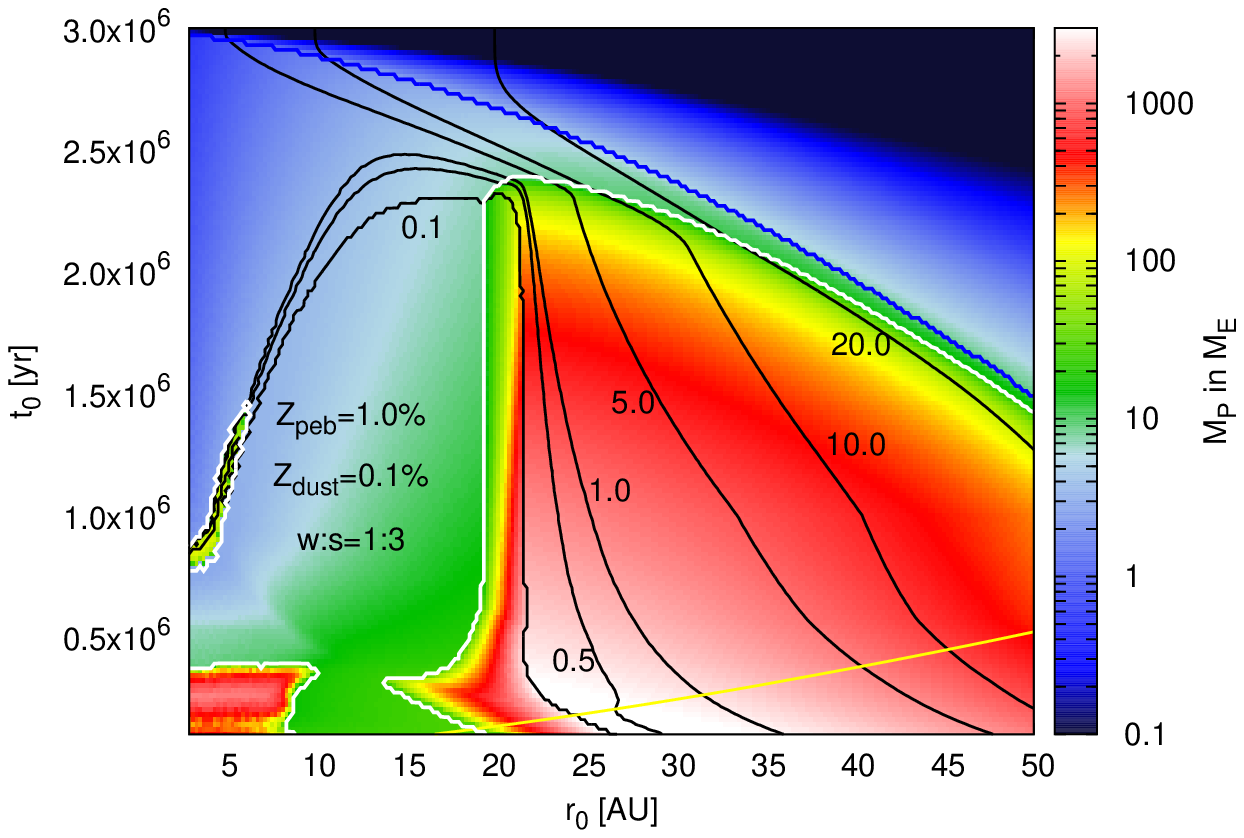}
 \caption{Final masses of planets ($M_{\rm P} = M_{\rm c}+M_{\rm env}$) as a function of initial radius $r_{\rm 0}$ and initial time $t_{\rm 0}$ in discs with a metallicity of dust  $Z_{\rm dust}=0.1\%$. The top plot features a water-to-silicate ratio of 3:1, while the bottom plot features a water-to-silicate ratio of 1:3. The black, blue, yellow  and white lines have the same meaning as in Fig.~\ref{fig:Z010M05}.  Planets can outgrow the region of outward migration and drift inwards at lower masses before they reach runaway gas accretion, which allows for a larger parameter space for the formation of low mass planets in contrast to high $Z_{\rm dust}$ discs (Fig.~\ref{fig:Z010M05}).
   \label{fig:Z010M01}
   }
\end{figure}

In Fig.~\ref{fig:Z010M01} we show the final planetary masses and orbital distances of planetary seeds that are put into a disc at a starting time $t_0$ and a starting orbital distance $r_0$, where the disc has $Z_{\rm dust}=0.1\%$. 

There are two main differences in the general outcomes compared to the discs with $Z_{\rm dust}=0.5\%$, i) gas giants can be formed farther out in the disc and at later times and ii) planetary seeds placed at $r_0 <20$ AU and formed at $t_0>500$ kyr only grow to several Earth masses and end up in the inner disc ($r_f<5$ AU) unless they form very late ($t_0>2.5$ Myr). The reason why gas giants can be formed farther out in the disc is related to the smaller $H/r$ in discs with $Z_{\rm dust}=0.1\%$ (Fig.~\ref{fig:Hrdisc}). A smaller $H/r$ reduces the pebble scale height, which allows the transition from 3D to 2D pebble accretion at a lower planetary mass, hence increasing the growth rate. Additionally to that, a smaller $H/r$ also increases the Stokes number $\tau_{\rm f}$ of the pebbles, which in turn increases the growth rate. However, a smaller $H/r$ reduces the pebble isolation mass (eq.~\ref{eq:Misolation}), so that planetary cores are smaller in the $Z_{\rm dust}=0.1\%$ case compared to the $Z_{\rm dust}=0.5\%$ case, but nevertheless they are big enough to attract a gaseous envelope within the lifetime of the disc.

In the inner disc, the aspect ratio $H/r$ is also smaller, which therefore results in a very low pebble isolation mass and hence core mass. The cores then need a very long time to attract a gaseous envelope. However, even before runaway gas accretion can start at $M_{\rm env} \sim M_{\rm core}$ the cores are too massive to be contained in the regions of outward migration and migrate into the inner disc for all water-to-silicate ratios.

As in the simulations with $Z_{\rm dust}=0.5\%$, there are also differences in the resulting planetary masses and orbital distances related to the water-to-silicate ratio in the disc. In the case of the high water content, a larger region of $r_0$-$t_0$ parameter space allows for the formation of small planets at $0.1$ AU $<r_{\rm f} <0.5$ AU. This region shrinks as the water content in the disc decreases. The planets in this region have a larger core and total mass for discs with larger water content, because the pebble isolation mass is higher in those discs. In addition to that, planets need a higher mass to outgrow the region of outward migration in discs with high water content to drift into the inner disc. As the growth time is longer, the planets migrate less when they outgrow the region of outward migration until the disc disperses, so they do not migrate all the way close to the star with $r_{\rm f}<0.1$ AU. Additionally, a small region in parameter space ($r_0\sim 3-10$ AU and $t_0 \sim 700 - 1500$ kyr) exists where the planetary seeds grow to become gas giants with masses around Saturn's. This region of parameter space also shrinks with decreasing water content in the disc. 

In the case of a low water-to-silicate ratio, the planets need to grow less to outgrow the region of outward migration. Therefore they have longer time to migrate until disc dispersal - thus a smaller region of parameter space allows for the formation of planets with $0.1$ AU < $r_{\rm f} < 0.5$ AU, compared to discs with a high water content.

\section{Discussion}
\label{sec:discuss}

\subsection{Disc structure and planet migration}

The structure of the disc is determined in an equilibrium between heating and cooling, where the cooling function depends on the opacity and thus on the chemical composition of the disc. By changing the water-to-silicate ratio in the disc, the bump in $H/r$ at the water ice line created by the opacity transition there changes as well. Reducing the water content in the disc reduces the bump in $H/r$ at the ice line. For very low $\dot{M}$ values, the bump at the ice line is nearly not visible any more (Fig.~\ref{fig:Hrdisc}). However, for all water-to-silicate ratios the aspect ratio in the inner disc is very small, allowing only a very low pebble isolation mass and thus the formation of small planets.

In turn the local radial gradients in surface density, temperature and entropy determine the migration of planets. In particular the regions of outward migration are greatly influenced by the water-to-silicate ratio in the disc (Fig.~\ref{fig:Migcont}). A low water fraction in the disc only allows planets with small range of masses to migrate outwards compared to discs with a large water fraction. Even though the difference in the migration map seem small, they have important consequences for the formation of planets.

As the core masses are very low in the inner regions of the protoplanetary disc (since $H/r$ is low), the contraction of a gaseous envelope takes a very long time, prolonging the time before runaway gas accretion can start. In the case of the water-to-silicate ratio of 1:3, the planets outgrow the region of outward migration, while their core mass $M_{\rm core}$ is still higher than their envelope mass $M_{\rm env}$. This means that these planets can drift towards the inner regions of the disc without becoming gas giants.

On the other hand, the region of outward migration in a disc with a water-to-silicate ratio of 3:1 can accommodate planets of higher masses. But as the pebble isolation mass is still very low, planets in discs with a large water-to-silicate fraction can only outgrow the region of outward migration if they are already in a runaway gas accretion phase. Counterintuitively this means that water-rich super-Earth planets should be more common around stars that have a low water-to-silicate fraction.

If the overall dust content is low ($Z_{\rm dust}=0.1\%$), planets can outgrow the region of outward migration without becoming gas giants and migrate into the inner disc for all water-to-silicate ratios. But as soon as the planet reaches the inner edge of the disc with $r_{\rm f}<0.1$ AU, the evolution of the planet is stopped in our model, which means the planet could in principle change its mass until the disc disperses. This is not followed in our model due to the uncertainties of planet growth close to the inner edge of the disc. This means that if discs have a low dust content towards the end of their lifetime, predictions regarding the abundance of super-Earth planets as a function of the water fraction in the disc are not entirely certain. 

However, recent simulations including photoevaporation have shown that the dust to gas ratio increases in the late stages of the disc evolution \citep{2015ApJ...804...29G}. This means that the case of high dust metallicity may be more relevant, implying that water-rich super-Earths should be more common around stars that have a low water-to-silicate fraction.

\subsection{Masses of the planetary cores}

The masses of solids that each planetary seed can accumulate by pebble accretion are mainly determined by the pebble isolation mass, which in turn is determined by the disc structure (eq.~\ref{eq:Misolation}). When the pebble isolation mass is reached, pebble accretion stops.

\begin{figure}
 \centering
 \includegraphics[scale=0.7]{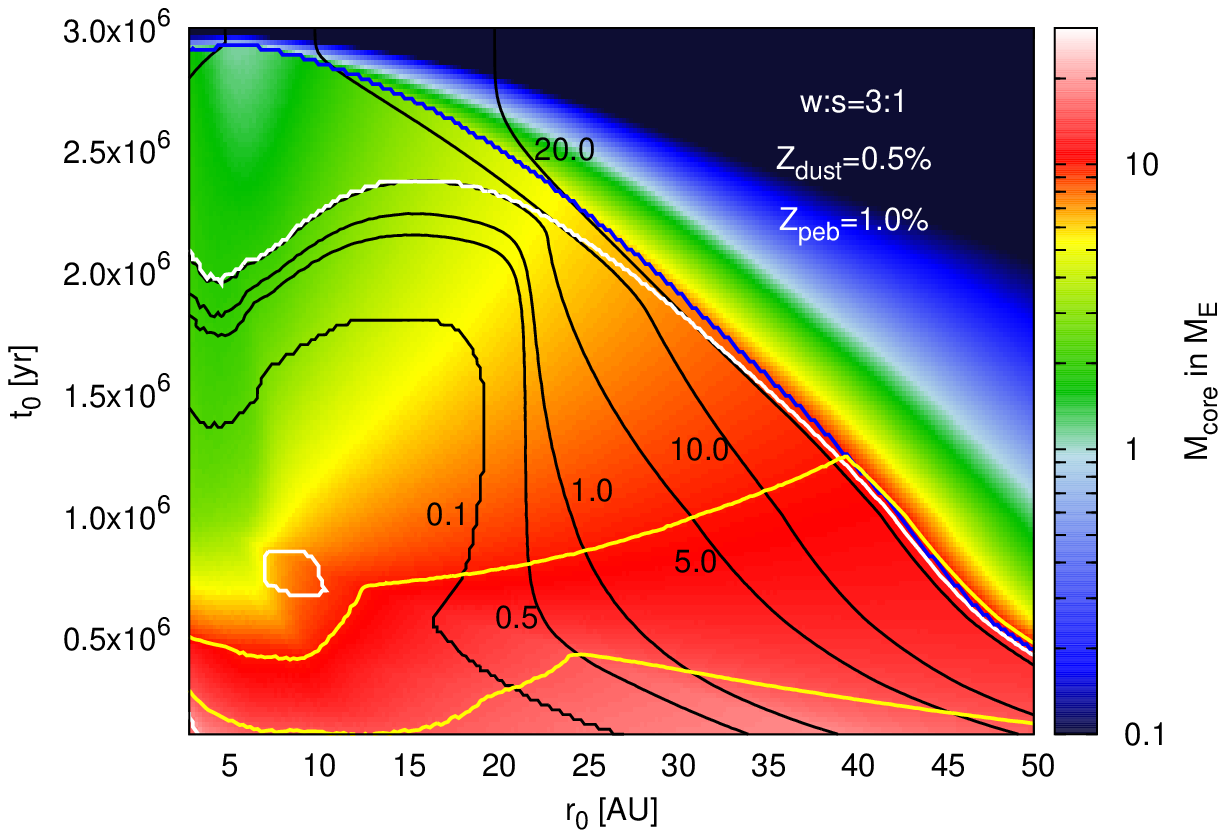}
 \includegraphics[scale=0.7]{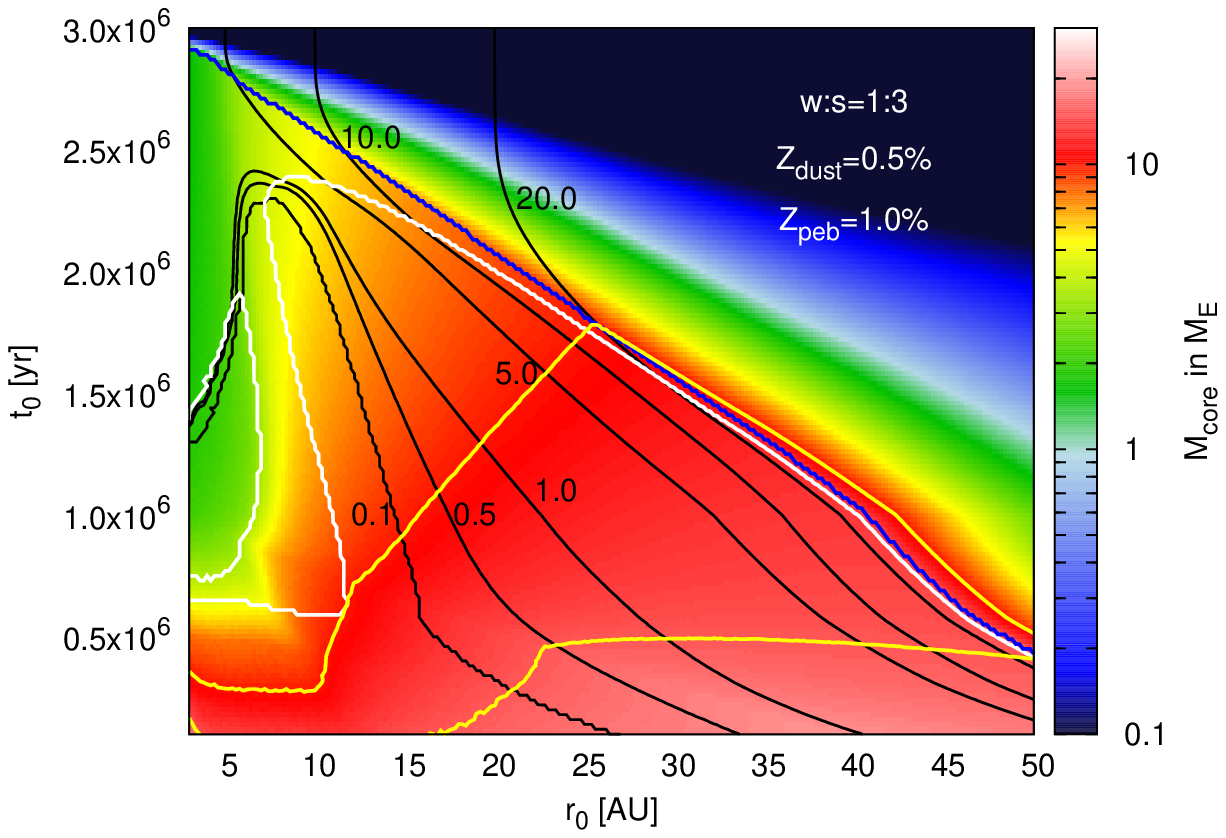}
 \caption{Final core masses of planets that start with planetary seeds at $r_{\rm 0}$ and $t_0$ for the disc with $Z_{\rm dust}=0.5\%$. The top features a water-to-silicate ratio of 3:1, while the bottom plot features a water-to-silicate ratio of 1:3. The blue and white lines have the same meaning as in Fig.~\ref{fig:Z010M05}. Planets that are below the top yellow line have a core mass of at least $10$ Earth masses, planets that are below the second yellow line from top have core masses of at least $20$ Earth masses. The final core mass is determined by the pebble isolation mass, which increases with orbital distances, because $H/r$ increases. However, at the outer edges of the disc, pebble accretion is not that efficient any more, so that pebble isolation mass is not reached, leaving the final core mass lower than at smaller orbital distances.
   \label{fig:Coremass}
   }
\end{figure}

\begin{figure}
 \centering
 \includegraphics[scale=0.7]{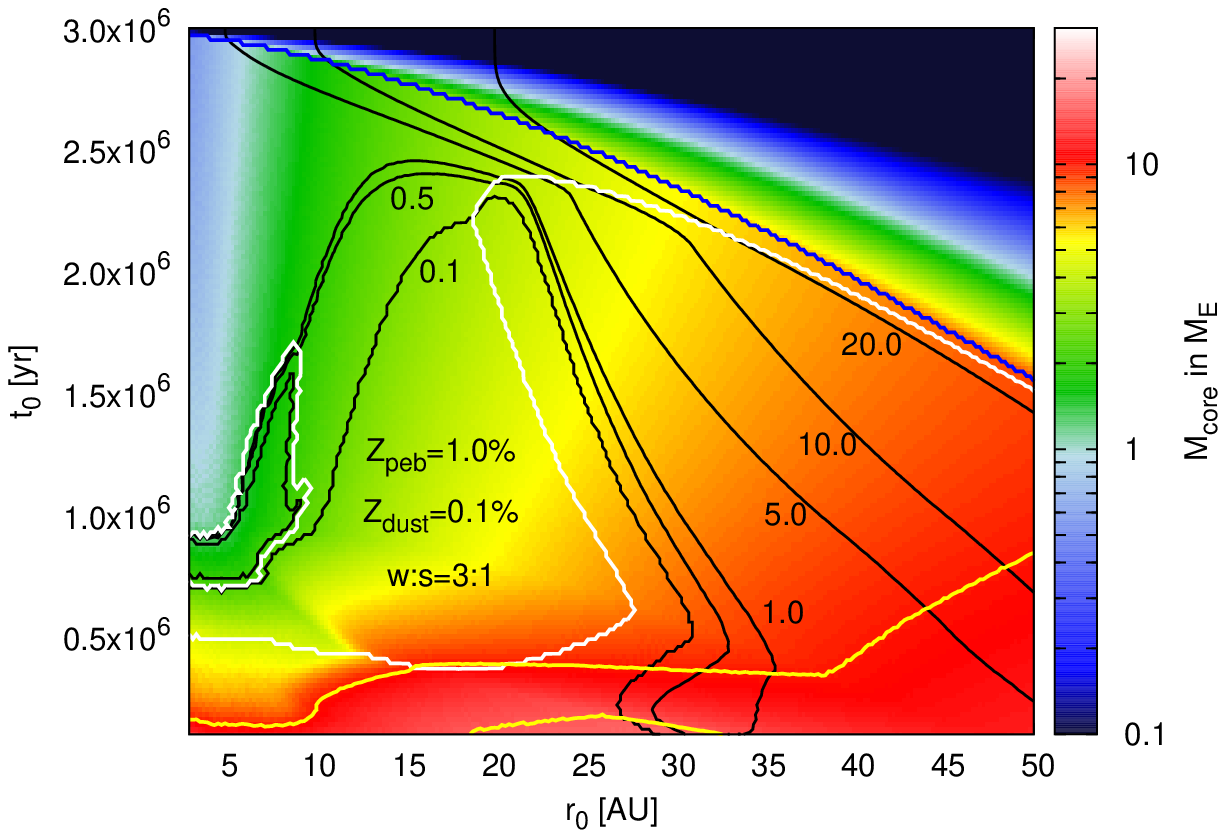}
 \includegraphics[scale=0.7]{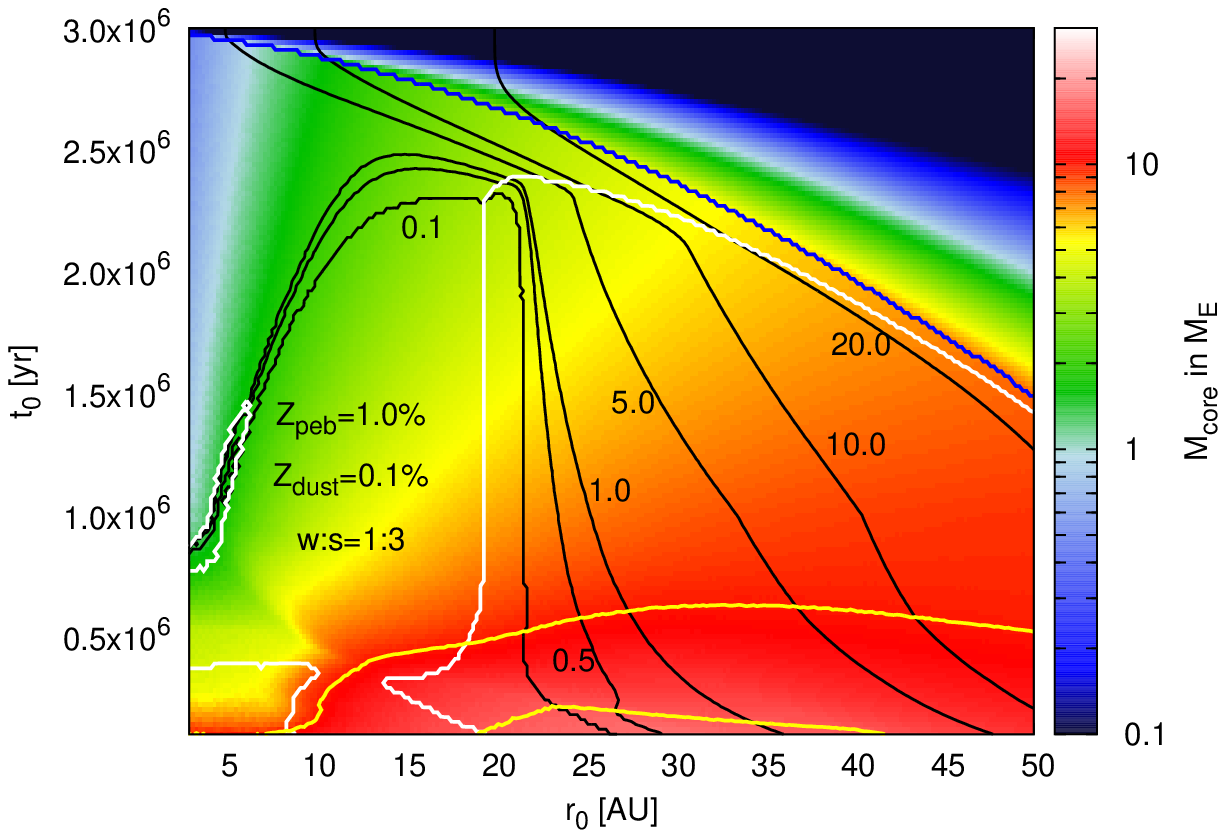}
 \caption{Final core masses of planets that start with planetary seeds at $r_{\rm 0}$ and $t_0$ for the disc with $Z_{\rm dust}=0.1\%$. The top features a water-to-silicate ratio of 3:1, the bottom plot a water-to-silicate ratio of 1:3. The coloured lines have the same meaning as in Fig.~\ref{fig:Coremass}. The final core masses are lower compared to the $Z_{\rm dust}=0.5\%$ disc, because of the lower pebble isolation mass in the outer disc. This also means that planetary seeds can reach pebble isolation mass at larger orbital distances and later times compared to seeds in the $Z_{\rm dust}=0.5\%$.
   \label{fig:Coremass01}
   }
\end{figure}

Because a larger $H/r$ allows a higher pebble isolation mass (eq.~\ref{eq:Misolation}), planetary seeds that form early in the disc can form larger planetary cores. Additionally this effect is enhanced if the disc contains a larger amount of micrometre-sized dust, as this increases $H/r$ in the disc (Fig.~\ref{fig:Hrdisc}). The final planetary core mass for planets in discs with high and low dust content for the water-to-silicate ratios of 3:1 and 1:3 are shown in Fig.~\ref{fig:Coremass} and Fig.~\ref{fig:Coremass01}. Clearly a later starting time $t_0$ results in smaller planetary cores.

Generally higher core masses are achieved if the planetary seed forms early and in a silicate rich disc. On the other hand, the formation of gas giants at large orbital distances seems easier in water rich discs, because the cores can grow faster (smaller $H/r$ leads to a smaller pebble scale height, allowing the transition to 2D accretion earlier). The final core mass in the water rich disc is generally lower, resulting in a longer phase of gas envelope contraction. However, this longer envelope contraction phase is overcompensated by the faster growth of the planetary core, which allows the formation of gas giant planets in the water rich disc at larger orbital distances and late formation times.

However, the masses of the planetary cores are about a factor of $1.5-2$ too small compared to the cores of the giant planets in our solar system ($\sim 15-30$ ${\rm M}_{\rm Earth}$), independently of the water-to-silicate fraction. A possible solution to this discrepancy could be that protoplanetary discs in their late stages are more dusty than the normally invoked dust to gas ratio of $0.01$. Alternatively, the pebble isolation mass inferred from isothermal 3D simulations in \citet{2014A&A...572A..35L} may miss a dependence of the isolation mass on the local slope of the pressure gradient. We plan to address this possibility in a future publication.

Another alternative to allow larger planetary cores is a different evolution of the protoplanetary disc. The reason why the core masses are quite low (Fig.~\ref{fig:Coremass} and Fig.~\ref{fig:Coremass01}) is that the low aspect ratio decreases in time \citep{2015A&A...575A..28B}. If the accretion rate $\dot{M}$ of the disc would decrease at a slower rate, then a larger $H/r$ could be maintained for a longer time, allowing for higher core masses. However, in order to avoid too far inward migration of the formed planets, the disc lifetime has to be roughly the same. This implies that the disc must disperse at a higher accretion rate compared to the simulations shown here, where the final accretion rate at $3$ Myr is $\dot{M}=2 \times 10^{-9}$M$_\odot/$yr. Given the ranges of accretion rates at which photoevaporation can be efficient, ($\dot{M}_{\rm evap} \sim \dot{M}=2-10 \times 10^{-9}$M$_\odot/$yr) see \citet{2013arXiv1311.1819A}, it is possible that the solar protoplanetary disc had a large accretion rate that could explain the higher planetary cores masses of the giant planets in the solar system.

The masses of the cores of small planets that drift into the inner disc are only up to a few Earth masses. As those planets have also reached pebble isolation mass, but not runaway gas accretion yet, they contain an envelope of gas, which can contain up to $50\%$ of the planet's total mass. The density of observed exoplanets peaks on average at $\sim 5$ ${\rm M}_{\rm Earth}$, indicating planets that are mostly rocky, in contrast to more massive exoplanets that contain a gaseous envelope \citep{2014ApJ...783L...6W}. This is in contrast to our simulation, where the planets drifting into the inner disc contain gaseous envelopes, which reduces their mean density. 

However, in our simulations the effects of the accretion of planetesimals is ignored. Planetesimal accretion can not only increase the accretion rate in the inner disc to some level, but also the constant bombardment of planetesimals can also prolong and even hinder the accretion of a gaseous envelope \citep{1996Icar..124...62P}, allowing for a higher mean density in agreement with observations.

Additionally, models of \citet{2015arXiv150905772L} have shown that the gaseous envelope of a super-Earth can be stripped away by impacts after the gas disc dispersed. In our formation model, where cores of $2-3$ ${\rm M}_{\rm E}$ with envelopes of $2-2.5$ ${\rm M}_{\rm E}$ drift into the inner disc, collisions are not taken into account. But one can imagine that collisions in the inner system take place and thus give rise to the formation of dense rocky planets without gaseous envelopes. On the other hand our model explains naturally the observed planet population that have a gaseous envelope. Those planets did not undergo collisions after the gas disc dispersed, allowing them to keep their gaseous envelope.

The masses of planetary cores growing via pebble accretion depends crucially on the disc's aspect ratio. A lower aspect ratio allows a faster growth (eq.~\ref{eq:Hpebble}), but results in a lower final mass of the planet (eq.~\ref{eq:Misolation}), while a larger aspect ratio slows down the initial growth, but allows a higher final planetary mass. In our simulations we study the influence of the change of the disc structure due to the water-to-silicate ratio on planet formation, but we kept the viscosity of the disc fixed and uniform at all orbital distances. Studying all different disc configurations is unfortunately computationally to expensive for a study like this, but we want to comment on the possible outcomes of a study like this.

\citet{2014A&A...570A..75B} studied the influence of a dead zone (ranging from $\sim 1$ to $\sim 10$ AU at high accretion rates) on the thermal structure of protoplanetary discs. As most of the mass can be carried by the upper layers, a reduction of viscosity in the mid-plane of the disc does not lead to an equally large increase in surface density there. As a consequence the disc is colder compared to a disc without a dead zone. This leads to a faster initial pebble accretion rate, so that the cores can form faster. However, the pebble isolation mass is lower as well, so that the final core mass is lower than in our nominal model. Nevertheless the aspect ratio is probably high enough so that the cores can grow to a few Earth masses, indicating that they can form gas giants in the end. 

As the disc evolves in time, the dead zone shrinks and finally disappears \citep{2013ApJ...765..114D, 2014A&A...570A..75B}, which means that planet formation at the late stages of the disc evolution is independent of the assumption of dead-zone in the disc's mid-plane. More importantly, our simulations show that at early stages mostly gas giants are formed, which is at odds with observations showing the abundance of super-Earths around solar-type stars. We therefore already suggested in \citet{2015A&A...582A.112B} that the formation of the planetary seeds at pebble transition mass starts late. This then indicates that our model is independent on the assumption of a planetary dead zone.

\subsection{Solar system}

In a recent study \citet{2015Natur.524..322L} incorporated accretion of pebbles onto planetary embryos in an N-body code that additionally follows the dynamical, collisional, and accretional evolution of a large number of planetesimals. If the pebbles form slowly enough, so that the planetesimals can gravitationally interact with one another, the larger planetesimals scatter their smaller siblings out of the disc of pebbles thus hindering the growths of the small planetesimals by pebble accretion. In their study this typically leads to the formation of one to four planetary cores of a up to $10$ ${\rm M}_{\rm Earth}$ between $5$ and $15$ AU. These planetary cores can then accrete gas, so they could resemble the structure of the outer solar system at the beginning of the Nice model \citep{2005Natur.435..459T}.

However, in \citet{2015Natur.524..322L} 
\begin{itemize}
 \item[1)] the protoplanetary disc structure is represented by a simple power law, which does not represent the structure of the protoplanetary disc in the inner parts (see Fig.~\ref{fig:Hrdisc})
 \item[2)] the accretion of gas onto the planet after it has reached pebble isolation mass is not modelled
 \item[3)] the migration of planetary embryos is ignored, even though it strongly influence the planetary system \citep{2015A&A...582A.112B} 
\end{itemize}
In our planet growth model, we take into account all of these physical effects, however we are missing the N-body approach with interacting planetesimals. But as \citet{2015Natur.524..322L}  have shown that only a small number of few planetary cores dominate in mass, we can use our planetary formation code to reproduce the initial conditions of the Nice model including an evolving protoplanetary disc structure, gas accretion and planetary migration.

\begin{figure}
 \centering
 \includegraphics[scale=0.7]{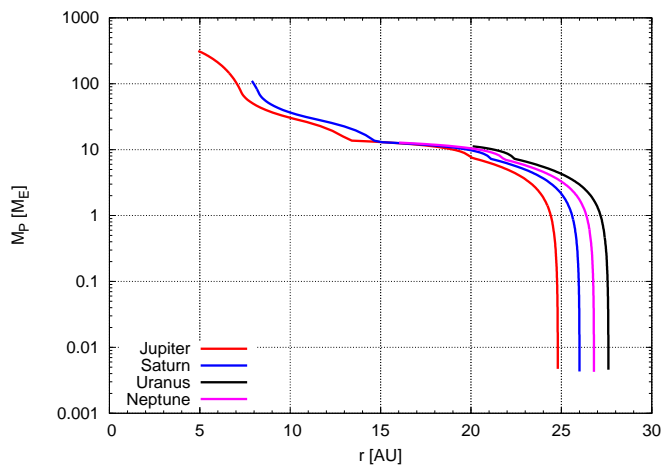}
 \caption{Evolution of the total planetary masses as a function of orbital distance for a solar-system analogue with the giant planets ending up in a configuration similar to the initial conditions of the Nice model. Jupiter starts at $t_{\rm 0}=1.68$ Myr, Saturn and $t_{\rm 0}=1.92$ Myr, Neptune and Uranus both start at $t_{\rm 0}=2.04$ Myr. We display Neptune inside of Uranus here, as they often switch places during the Nice model \citep{2005Natur.435..459T}. We have used here $Z_{\rm dust} = 0.5\%$ and a metallicity of pebbles of $Z=1.0\%$ where the disc has a water-to-silicate ratio of 3:1. The mass of the cores for all giant planets is $\sim 7$ ${\rm M}_{\rm Earth}$.
   \label{fig:Solarsystem}
   }
\end{figure}

In Fig.~\ref{fig:Solarsystem} we show the growth tracks of the four gas giants in our solar system in a disc with $Z_{\rm dust}=0.5\%$, $Z_{\rm peb}=1.0\%$  and a water-to-silicate ratio of 3:1. We can very nicely reproduce the orbital distance and planetary masses for Jupiter and Saturn within the limits of the Nice model. Also our analogues for Neptune and Uranus fit within those parameters, but they are slightly too small. However, the formation of the ice giants could also be aided by giant impacts of planetary embryos that form outside of Jupiter and Saturns orbit \citep{2015arXiv150603029I}.

For water-to-silicate ratios of 1:1 and 1:3, the formation of the exact initial conditions of the Nice model is not possible, as Jupiter is slightly too small \citep{2015A&A...582A.112B}. The main problem here is not the formation of a Jupiter mass planet at $\sim 5$ AU, but to have at the same time a Saturn mass planet at $\sim 8$ AU, as we do not want the orbits of planets to cross at any time during their evolution.

In the case of $Z_{\rm dust}=0.1\%$, the initial configuration of the Nice model can be reproduced for all different water-to-silicate ratios in the disc. The main difference is that now the outermost planet has to be formed around $\sim 30$ AU, which is a slightly larger orbital distances than for discs with a larger dust content. The reason for that is the smaller $H/r$ in discs with a low dust content that reduces the pebble isolation mass and thus determines the mass of the planetary core.

In our simulations, the planetary seeds for the giant planets form around $25$ AU and then migrate inwards closer to their final orbital positions. At the late formation times of the planetary seeds, the disc is already very cold ($T<25$ K) in the outer parts of the disc. At these temperatures, CO can freeze out and is also detectable in protoplanetary discs around sun like stars \citep{2013Sci...341..630Q}. In particular \citet{2013Sci...341..630Q} observed the CO ice line at $\sim 30$ AU matching the temperature profile in our simulations. \citet{2014ApJ...793....9A} showed that the chemical composition of the giant planets in our solar system can be naturally explained if their formation starts at the CO ice line. This means our model can not only explain the formation of giant planets in the solar system, but also their composition.

\subsection{Correlation of planetary mass with host star metallicity}

The observations of exoplanets have revealed that giant planets are more common around stars that have an increased metallicity compared to the solar value \citep{2005ApJ...622.1102F, 2014Natur.509..593B, 2015A&A...580L..13S}. In particular the study of \citet{2014Natur.509..593B} showed that giant planets (with $R_{\rm P} > 4 R_{\rm Earth}$) are rarely found around stars with a metallicity of $[{\rm Z}/{\rm H}]\leq-0.2$. This was explained simply by the fact that not enough building blocks in solids were available to form planetary cores that are big enough to accrete gas to become gas giants. However, the very efficient pebble accretion mechanism for building cores could provide a fast enough growth to form big enough cores that can accrete gas contradicting the observations.

\begin{figure}
 \centering
 \includegraphics[scale=0.7]{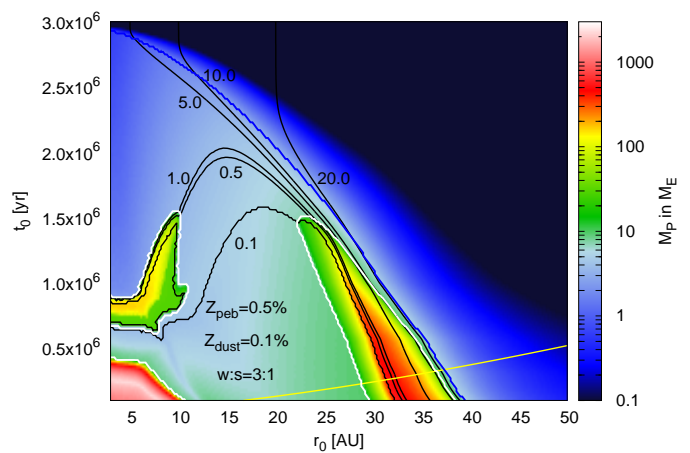}
 \caption{Final masses of planets that start with planetary seeds at $r_{\rm 0}$ and $t_0$ in a disc with $Z_{\rm dust}=0.1\%$ and a metallicity in pebbles of $Z_{\rm peb} = 0.5\%$. The blue, white and yellow lines have the same meaning as in Fig.~\ref{fig:Z010M05} and the disc features a water-to-silicate ratio of 3:1.
   \label{fig:Z005M01}
   }
\end{figure}

The study of \citet{2014A&A...562A..71B} investigated the metallicity and composition of dwarf stars close to our sun. They find that stars with a metallicity of $[{\rm Fe}/{\rm H}]\sim-0.2$ are enriched in oxygen compared to silicate by $1.4:1$. This could imply that the discs around those stars were enriched by water compared to silicate as well. An enrichment in water compared to silicate in the disc actually favours giant planet formation (see Fig.~\ref{fig:Z010M05}), so we present in Fig.~\ref{fig:Z005M01} the final planetary masses for planetary seeds that were placed in a disc with $Z_{\rm peb} = 0.5\%$, $Z_{\rm dust} = 0.1\%$ and a water-to-silicate ration of $3:1$ at an initial time $t_0$ and with initial orbital distances $r_0$.

As can be seen in Fig.~\ref{fig:Z005M01} the parameter space that allows for the formation of gas giants like Jupiter is very small. The formation of giant planets is only possible in the inner parts of the disc at early times and in the very out parts of the disc. In the inner disc at early times the disc is hot enough, so that planets can reach a high pebble isolation mass, reducing the contraction phase of the gaseous envelope and therefore allowing fast gas accretion before the planet's migration stops at the inner disc edge. The planetary seeds in the outer disc can reach a higher isolation mass, because the disc is flared, and are thus able to grow to become gas giants, however only if they form early enough to leave enough time to form the planetary core.

\section{Summary}
\label{sec:summary}

In this paper we studied the influence of a different water-to-silicate ratio and metallicity in dust grains $Z_{\rm dust}$ on the structure of protoplanetary discs and the formation and migration of planets. A larger amount of dust particles results in a hotter disc, independent on the exact chemical composition of the disc. A change in the water fraction changes the opacity, because the absorption and re-emission properties of the dust particles changes dependent on the chemical composition of the dust particles. This then changes the cooling function of the disc and hence the temperature profile of the disc. A change of the temperature profile then directly changes the scale height of the disc ($T\propto H^2$) and thus the viscosity inside the disc. In turn, this changes the surface density profile, because we drive our simulations to a constant accretion rate $\dot{M}$ at each radial distance (eq.~\ref{eq:Mdot}).

These changes in the disc structure have direct consequences on the migration of planets inside the disc, because the migration direction and speed depends on the gradients of surface density, temperature and entropy. In particular discs that have a larger water fraction can harbour slightly more massive planets in the regions of outward migration at the late stages of the disc evolution compared to discs with a small water fraction. This has important consequences for the formation and evolution of planets in those discs, because planets with lower masses can outgrow the region of outward migration easier and migrate inwards easier in discs with a smaller water content. This implies that planets with lower masses can only migrate to orbits close to the host star if the water content in the disc is low.

Planetary seeds that are embedded in the protoplanetary disc start to grow initially through the accretion of pebbles until they reach their pebble isolation mass \citep{2014A&A...572A..35L}, which is when gas accretion onto the planetary core can start. The growth rate of the core depends on the amount of pebbles in the disc and on the initial location $r_0$ and initial time $t_0$ when the planetary seeds is placed in the disc. This also influences the final mass of the core, because the pebble isolation mass is a strong function of the disc structure (eq.~\ref{eq:Misolation}), which evolves in time (Fig.~\ref{fig:Hrdisc}). In particular the aspect ratio of the disc decreases in time, resulting in a lower pebble isolation mass and hence a lower final core mass.

After reaching pebble isolation mass, the planetary core can start to accrete gas. However before runaway gas accretion can start, the gaseous envelope needs to contract until $M_{\rm env} \sim M_{\rm core}$. This contraction phase is shorter for larger planetary cores \citep{2014ApJ...786...21P}, meaning that small cores (around $2$ ${\rm M}_{\rm Earth}$) take a very long time to contract their envelope.

Discs with a larger dust content ($Z_{\rm dust} = 0.5\%$) and a larger water fraction allow the formation of giant planets at later times and farther away from their host star compared to discs with a small water fraction. On the other hand, a larger silicate fraction in the disc allows the migration of small ice planets and ice giants to hot and warm orbits close to the host star, which is not possible in the disc with a large water content. This is caused by the fact that lower mass planets can outgrow the region of outward migration and move inwards before they enter the runaway gas accretion phase and become gas giants.

If a lower dust content ($Z_{\rm dust} = 0.1\%$) is present in the disc, the amount of ice planets and ice giants that can drift to the inner disc is increased independent of the water-to-silicate mixture. However, photoevaporation can increase the dust-to-gas ratio in the late stages of the disc \citep{2015ApJ...804...29G}, indicating that the disc model with $Z_{\rm dust}=0.5\%$ may be the more common case in the late stages of disc evolution when the major core growth occurs.

As the disc evolves in time, the structure of the disc changes. In particular the aspect ratio of the disc decreases in time (with decreasing $\dot{M}$) and with it the pebble isolation mass. Therefore, the later a planetary seed forms, the lower its final core mass. But even for early formation of planetary seeds we find that the final core mass is a factor $1.5-2$ too low when compared to the ice and gas giants in our own solar system. This problem can be overcome if the disc has a larger $\dot{M}$ during its evolution, which results in a hotter disc and higher cores masses, and disperses at a larger value of $\dot{M}$. Photoevaporation at a relatively high ${\dot M}$ of $\dot{M}_{\rm evap} \sim \dot{M}=2-10 \times 10^{-9}$M$_\odot/$yr  can occur if driven by X-rays  \citep{2009ApJ...699.1639E, 2013arXiv1311.1819A}.

Studies of the composition of exoplanets reveal that low mass planets ($M_{\rm P}< 5$ ${\rm M}_{\rm E}$) are mostly rocky, while more massive planets contain a gaseous envelope \citep{2014ApJ...783L...6W}. The small planets formed in our simulations that then migrate into the inner disc have reached pebble isolation mass and therefore contain a gaseous envelope smaller than their core mass. Collisions between these planets can then strip them from their atmosphere after the gas disc dispersed \citep{2015arXiv150905772L}, allowing the formation of planets without extended atmospheres. On the other hand our model explains naturally planets with  ($M_{\rm P}> 5$ ${\rm M}_{\rm E}$) harbouring a gaseous envelope.

We have shown in this paper that the chemical composition (the water-to-silicate ratio) of the micrometre-sized dust grains in the protoplanetary disc influences the disc structure and with it the migration and formation pattern of planetary cores in those discs. In particular, a low water abundance relative to silicates allows icy cores to migrate to sub-AU orbits. Thus our results imply that some observed super-Earths in orbits close to their host star should contain a large proportion of water accreted outside of the ice line.

\begin{acknowledgements}

B.B.,\,and A.J.\,thank the Knut and Alice Wallenberg Foundation for their financial support. B.B.\, also thanks the Royal Physiographic Society for their financial support. A.J.\,was also supported by the Swedish Research Council (grant 2010-3710), the European Research Council (ERC Starting Grant 278675-PEBBLE2PLANET) and the Swedish Research Council (grant 2014-5775). B.B.,\,and A.J.\,thank also C. Battistini and T. Bensby for information concerning stellar abundances. Additionally we thank the referee for his/her comments that helped to improve the manuscript.

\end{acknowledgements}

\appendix
\section{Type-II migration}
\label{ap:migration}
\begin{figure}
 \centering
 \includegraphics[scale=0.7]{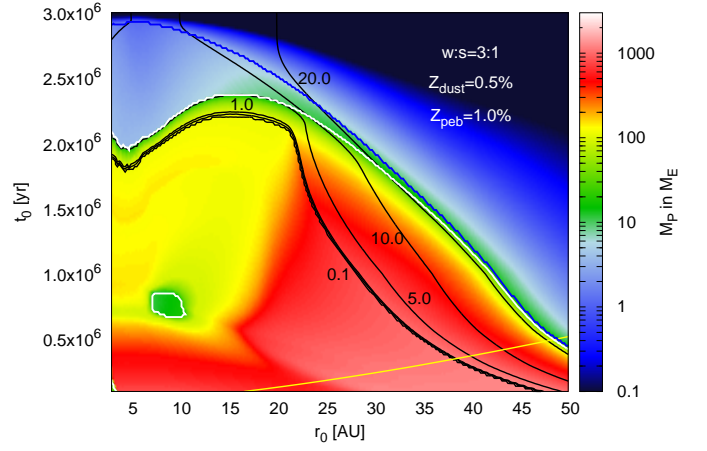}
 \caption{Final masses of planets that start with planetary seeds at $r_{\rm 0}$ and $t_0$ in a disc with $Z_{\rm dust}=0.5\%$ and a metallicity in pebbles of $Z_{\rm peb} = 1.0\%$. The blue, white and yellow lines have the same meaning as in Fig.~\ref{fig:Z010M05} and the disc features a water-to-silicate ratio of 3:1. However, in contrast to the previous simulations, we have turned off the reduction of the migration speed in type-II migration due to the planet's inertia.
   \label{fig:Z010M05noinert}
   }
\end{figure}

Growing planets migrate through the disc. In the initial stage, their migration rate is determined by type-I migration, but as soon as the planets become big enough to open gaps in the protoplanetary discs, their migration is described by type-II migration. Recent studies, however, have revealed that the type-II migration rate might not be related to the viscous evolution of the protoplanetary disc \citep{2015A&A...574A..52D}. Unfortunately, hydrodynamical studies of planet disc interactions have not progressed far enough to describe these effects in an updated type-II migration rate formula that can be used in a code like ours or in N-body codes. In our standard model, the type-II migration rate is described in eq.~\ref{eq:typeII}. In Fig.~\ref{fig:Z010M05noinert} we show the results of our planet formation model, if no reduction of the type-II migration rate due to the planet's inertia is taken into account.

Despite the fact that planetary migration is faster, our planet formation model with pebble accretion is still able to form giant planets. However, the final orbital distribution of planets is completely different. There are basically no planets stranded between $0.1$ and $1.0$ AU. All giant planets are now either inside of $0.1$ AU or outside of $1.0$ AU. The reason for this lies in the reduction of the type-II migration speed in eq.~\ref{eq:typeII}, where the speed of the planet is reduced more and more the closer the planet is to the central star. Additionally we now observe many more Saturn mass planets in the inner part of the disc compared to our standard model (Fig.~\ref{fig:Z010M05}).

Planetary seeds that form in the outer parts of the disc are not affected by the reduction of the type-II migration rate due to the inertia, because the outer disc is so massive that it dominates over the planetary mass. Late forming seeds are also unaffected by a change in the type-II migration rate, because they do not grow to become gap opening gas giants in the first place. 

Clearly, the type-II migration rate is important for the outcome of any planet formation model that wants to explain the formation of gas giants. But unfortunately many more hydrodynamical studies of planet-disc interactions have to be undertaken in order to constrain the type-II migration rate so that it can be incorporated in planet formation models.

\section{Envelope contraction}
\label{ap:envelope}
\begin{figure}
 \centering
 \includegraphics[scale=0.7]{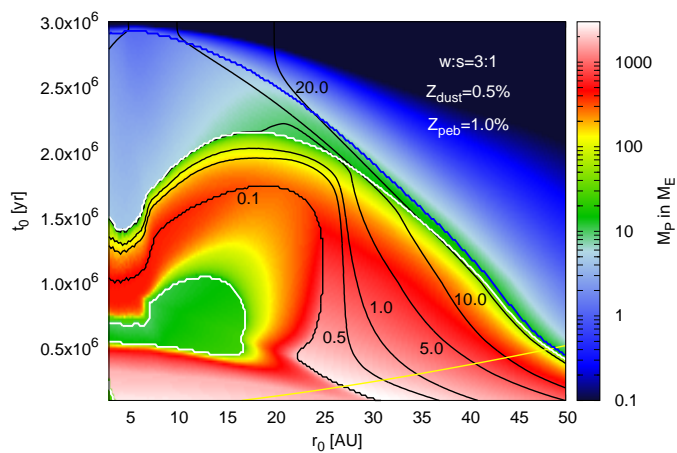}
 \caption{Final masses of planets that start with planetary seeds at $r_{\rm 0}$ and $t_0$ in a disc with $Z_{\rm dust}=0.5\%$ and a metallicity in pebbles of $Z_{\rm peb} = 1.0\%$. The blue, white and yellow lines have the same meaning as in Fig.~\ref{fig:Z010M05} and the disc features a water-to-silicate ratio of 3:1. However, in contrast to the previous simulations, the opacity for the contraction of the envelope is set to $\kappa_{\rm env} = 0.1{\rm cm}^2/{\rm g}$.
   \label{fig:Z010M05kappa01}
   }
\end{figure}

After the core reaches pebble isolation mass, it contracts a gaseous envelope. This contraction phase depends on the planetary mass and on the cooling rate of the planetary envelope. The cooling rate of the envelope is determined by the opacity $\kappa_{\rm env}$ inside it, which changes due to the amount and size distribution of grains. For the envelope contraction we follow \citet{2014ApJ...786...21P}, who give an accretion rate of
\begin{eqnarray}
\label{eq:Mdotenv}
 \dot{M}_{\rm gas} &= 0.000175 f^{-2} \left(\frac{\kappa_{\rm env}}{1{\rm cm}^2/{\rm g}}\right)^{-1} \left( \frac{\rho_{\rm c}}{5.5 {\rm g}/{\rm cm}^3} \right)^{-1/6} \left( \frac{M_{\rm c}}{{\rm M}_{\rm E}} \right)^{11/3} \nonumber \\ 
 &\left(\frac{M_{\rm env}}{1.0{\rm M}_{\rm E}}\right)^{-1} \left( \frac{T}{81 {\rm K}} \right)^{-0.5} \frac{{\rm M}_{\rm E}}{{\rm Myr}}
,\end{eqnarray}
where $f$ is a fudge factor to change the accretion rate in order to match numerical and analytical results, which is normally set to $f=0.2$ \citep{2014ApJ...786...21P}. The density of the core is set to $\rho_{\rm c}=5.5 {\rm g}/{\rm cm}^3$. Here we use $\kappa_{\rm env} = 0.05{\rm cm}^2/{\rm g}$, which is very similar to the values used in the study by \citet{2008Icar..194..368M}. In Fig.~\ref{fig:Z010M05kappa01} we have increased $\kappa_{\rm env}$ to $0.1{\rm cm}^2/{\rm g}$. The contraction time of the envelope is therefore twice as long.

The main differences are i) it is possible for ice giants to migrate all the way to the central star and ii) the formation of gas giants at later ages is hindered. The cores of the ice giant population form at a few AU from the host star and reach a moderate pebble isolation mass ($6-7$ Earth masses), which gives them also a moderate time for the contraction of the gaseous envelope. As the planets grow, they reach the region of outward migration and stay there during most of the envelope contraction phase, but eventually they become too massive and migrate towards the star before they reach runaway gas accretion. Formation at the same $r_0$, but at later times (larger $t_0$), results in lower core masses, which allows the planets to stay in the region of outward migration until gas runaway accretion is reached. These planets then become gas giants. The longer envelope contraction phase then also hinders planets forming at later times to reach runaway gas accretion, so that they stay outside of a few AU, trapped at the outer edge of the region of outward migration.

Clearly, the opacity inside the planetary envelope is very important for the outcome of planet formation studies. However, we assume that the opacity in the envelope is very small for two reasons:
\begin{itemize}
 \item[i)] As the planet reaches pebble isolation mass, it accelerates the gas outside of it is orbit to a super-Keplerian speed \citep{2014A&A...572A..35L}, which hinders particles to reach the planet. This means that the gas accreted by the planet is probably depleted in heavy elements.
 \item[ii)] The planet formation model via pebble accretion requires only a very small density of solids, while in classical core accretion scenarios a large surface density of planetesimals is invoked to form the core in the first place \citep{1996Icar..124...62P}. This larger amount of planetesimals also leads to larger collision rates of the gas accreting planet with planetesimals, which prolongs the contraction of the gaseous envelope. Additionally, simulations by \citet{2015Natur.524..322L} have shown that the smaller planetesimals are scattered out of the mid-plane of the disc, which should prevent them from being accreted by the dominant gas accreting planet. The combination of a low surface density of planetesimals and the effects of scattering of planetesimals result in less solids being accrete by the gas accreting planet, reducing the opacity in the envelope and therefore accelerating the contraction of the envelope.
\end{itemize}

\bibliographystyle{aa}
\bibliography{Stellar}
\end{document}